\documentclass[fleqn,usenatbib]{mnras}

\usepackage[T1]{fontenc}

\DeclareRobustCommand{\VAN}[3]{#2}
\let\VANthebibliography\thebibliography
\def\thebibliography{\DeclareRobustCommand{\VAN}[3]{##3}\VANthebibliography}

\hypersetup{linkcolor=blue,citecolor=blue,filecolor=cyan,urlcolor=magenta}
\usepackage{graphicx}   
\usepackage{amsmath}    
\usepackage{amssymb}    
\usepackage{newtxtext,newtxmath}
\usepackage{xcolor}
\usepackage{ulem}
\usepackage{threeparttable}
\usepackage{hyperref}

\newcommand{\lta}{\lower 2pt \hbox{$\, \buildrel {\scriptstyle <}\over {\scriptstyle \sim}\,$}}
\newcommand{\gta}{\lower 2pt \hbox{$\, \buildrel {\scriptstyle >}\over {\scriptstyle \sim}\,$}}
\definecolor{blazeorange}{rgb}{1.0, 0.4, 0.0}
\definecolor{seagreen}{rgb}{0.18, 0.55, 0.34}
\definecolor{rufous}{rgb}{0.66, 0.11, 0.03}
\definecolor{royalfuchsia}{rgb}{0.79, 0.17, 0.57}
\definecolor{scarlet}{rgb}{1.0, 0.13, 0.0}
\definecolor{royalpurple}{rgb}{0.47, 0.32, 0.66}

\title[FRB source size]{Constraining the FRB mechanism from scintillation in the host galaxy}

\author[Kumar et al.]{
Pawan Kumar$^1$\thanks{pk@astro.as.utexas.edu}, Paz Beniamini$^{2,3,4}$\thanks{pazb@openu.ac.il}, Om Gupta$^1$\thanks{omgupta@austin.utexas.edu} \& James M. Cordes$^5$\thanks{cordes@astro.cornell.edu} \\ \\
$^1$Department of Astronomy, University of Texas at Austin, Austin, TX 78712, USA\\
$^{2}$Department of Natural Sciences, The Open University of Israel, P.O Box 808, Ra'anana 4353701, Israel\\
$^{3}$Astrophysics Research Center of the Open university (ARCO), The Open University of Israel, P.O Box 808, Ra'anana 4353701, Israel\\
$^4$Department of Physics, The George Washington University, 725 21st Street NW, Washington, DC 20052, USA\\
$^{5}$Astronomy Department, Cornell University, Ithaca, NY 14853, USA
\\
}

\begin{document}
\label{firstpage}
\pagerange{\pageref{firstpage}--\pageref{lastpage}}
\maketitle

\begin{abstract}

Most FRB models can be divided into two groups based on the distance of the radio emission region from the central engine. 
The first group of models, the so-called `nearby' or magnetospheric models, invoke FRB emission at distances of 10$^9$ cm or less from the central engine, while the second `far-away' models involve emission from distances of 10$^{11}$ cm or greater. The lateral size for the emission region for the former class of models ($\lta$ 10$^7$ cm) is much smaller than the second class of models ($\gta 10^9$ cm). We propose that an interstellar scattering screen in the host galaxy is well-suited to differentiate between the two classes of models, particularly based on the level of modulations in the observed intensity with frequency, in the regime of strong diffractive scintillation. This is because the diffractive length scale for the host galaxy's ISM scattering screen is expected to lie between the transverse emission-region sizes for the `nearby' and the `far-away' class of models. Determining the strength of flux modulation caused by scintillation (scintillation modulation index) across the scintillation bandwidth ($\sim 1/2\pi\delta t_s$)
would provide a strong constraint on the FRB radiation mechanism when the scatter broadening ($\delta t_s$) is shown to be from the FRB host galaxy. The scaling of the scintillation bandwidth as $\sim \nu^{4.4}$ may make it easier to determine the modulation index at $\gta$ 1 GHz.

\end{abstract}




\section{Introduction}
\label{sec:intro}

Fast Radio Bursts (FRBs) are a class of highly energetic short-duration astrophysical transients which were discovered in 2007 \citep{Lorimer+07}. The vast majority of these events are extragalactic, with observed flux densities between a few mJy to a few hundred Jy and observed durations ranging from few $\mu$s (e.g., \citealt{Snelders2023, Hewitt2023}) to several ms. Numerous studies have tried to characterise the source properties and emission mechanism by probing wide ranges in frequency space, between 110 MHz \citep{Pleunis2021a} and 8 GHz \citep{Gajjar2018}. The FRB sources are suggested to be highly compact, with various lines of evidence pointing towards magnetars as the sources of (at least some) FRBs, including one repeating FRB source, FRB20200428, associated with well-studied Galactic magnetar \citep{STARE2020,CHIME2020}.
FRB radiation models fall into two broad categories: magnetospheric and far-away. The magnetospheric model suggests that the coherent radio waves are generated within the magnetosphere of a neutron star, while the far-away model suggests that the source is near or outside the light-cylinder.
The case for a magnetospheric origin of FRBs is supported by many theoretical arguments (e.g., \citealt{BK2020,wang2019,LKZ2020,wang+20,zhang-review2020,LBK2022,BK2023}) and some FRB data \citep{Nimmo2021, Nimmo2022, Zhang2023, Snelders2023}. \cite{CW2016}, based on an earlier work (\citealp{2004ApJ...612..375C}), have suggested that FRBs are composed of incoherent superposition of a large number of coherent pulses each of which are roughly of ns duration that are produced in the magnetosphere. This general picture seems consistent with the FRB data
\citep[e.g.][]{Nimmo2022}. Some concerns have been raised about the magnetospheric model's ability to produce FRB radiation due to the magnetar magnetosphere's opacity \citep{Beloborodov2021}. However, recent research shows that large amplitude, coherent, radio waves can escape under certain conditions \citep{Qu2022}. Another issue with the magnetospheric model is the  possible difficulty of the central engine operating for $\gta 10$ms to produce the longest-duration FRBs \citep{Metzger+19}. This is because the natural timescale for any disturbance in the {neutron star (NS) crust is of the order of a few ms, which is the duration over which the magnetospheric disturbance should last. The far-away class of models for FRBs also has many drawbacks. The most severe among these are the short time variability compared with burst duration and narrow band spectra for many bursts \citep{BK2020}, large induced-Compton optical depth in the upstream medium \citep{KumarLu-IC-scat-2020}, and the discovery of a recent highly periodic series of peaks in a long duration non repeating FRB \citep{subsecPCHIME,BK2023}. The debate about the origin of FRB coherent radiation is ongoing and is expected to be resolved by additional data or reanalysis of existing data. The analysis we propose in this work could help settle the debate by providing a constraint on the source region size.
While much of our discussion focuses on the case of magnetars as the central engine for FRBs, we emphasize that the basic idea presented in this work does not rely on the specific association of FRBs with magnetars. It could be applied to any FRB model because the primary consideration of the method developed in this paper is the lateral size of the emitting region.

According to magnetospheric models, the source size of FRBs is small, $\lta 10^7$cm (cf. \citealt{Kumar+17,KumarBosnjak2020}). On the other hand, the far-away models propose that the source size is around $R/\gamma\gta 10^{9}$cm \citep{Lyubarsky14,Metzger+17,Beloborodov17,Metzger+19,Beloborodov19,Margalit+20}, and this is determined by the distance $R$ from the neutron star where the radiation is produced by a relativistic outflow moving at the Lorentz factor $\gamma$.

The turbulent plasma between the FRB source and the radio telescope scatters the FRB pulse and limits the observed FRB spectrum's coherence bandwidth.
Observations of FRBs show evidence of scintillation originating from scattering screens residing both within the host galaxy and within the Milky Way \citep{Farah2018, Ocker2022b, Sammons2023}. Scattering in FRB host galaxies has only been observed as the asymmetric pulse broadening produced by propagation along multiple paths. From the broadening time, we can infer the scintillation bandwidth of any intensity scintillations. For measured broadening times that are typically $\sim1$ ms or longer, the scintillation bandwidth is too small to measure at $\sim$GHz frequencies. However, it is worth noting that many FRBs do not exhibit pulse broadening. In such cases, the upper bounds on the broadening suggest that scintillation bandwidths could be large enough to be measurable. In addition, the strong frequency dependence of scintillation bandwidths $\propto\nu^{4.4}$ suggests that higher-frequency observations can yield direct measurements of scintillations in the FRB host galaxy.

Evidence for scattering of radio waves within the Milky Way (MW) has come from measuring intensity fluctuations over frequency scales greater than the scintillation bandwidth for all FRBs, with the exception of 20180916B at low frequencies where temporal broadening due to scattering in the MW-ISM was observed \citep[][]{2021Natur.596..505P, Gopinath2023}.

Whenever there is a  significant separation of scales between the scintillation bandwidths for the MW and the FRB-host scattering screens, the two can be separately identified in the data and used for probing different aspects of FRB physics.

A plasma screen acts like a telescope with high angular resolution that can help determine the source size and hence narrow down the radiation mechanism. The diffractive scale for the host galaxy plasma screen, as projected on the source, is generally much smaller than that for the MW-ISM screen. Therefore, the former can probe FRB sources to a smaller transverse scale and thereby help determine the radiation mechanism as different classes of models suggest widely different source sizes. When FRB radiation is produced within a small region of the magnetosphere, the amplitude of flux variations is of the order of unity over a frequency scale a few times larger than the scintillation bandwidth whereas it is small for an extended source like those suggested by far-away models. This offers a way to determine the FRB radiation process or at least narrow it down using FRB data. The challenge for observers is to identify FRB host galaxy scintillation and measure the flux variation amplitude across frequency scales greater than the scintillation bandwidth.

The idea that scintillation can be used to constrain the source size for radio pulsars has been extensively developed by many people over several decades, including, for example, \cite{1970PhDT.......113L,cwb83,Gwinn1997,jmc2000,Gwinn2012}, \cite{Johnson2012}, \cite{Lin2023}, and references therein. There are both similarities and many differences between the idea presented here and the work on pulsars. 
One of the primary distinctions in our current work is our focus on distinguishing between near-field and far-away models for FRB radio emission, which predict significantly different source sizes. The diffractive scale for strong scintillation falls between these two sizes.
Previous work on using scintillation to resolve pulsar magnetospheres has investigated intensity variations in both time and frequency, as described in papers cited above 
(see also \citet{Gwinn1997, Gwinn2012}).
Our current work centers on the variability of the FRB spectrum with frequency in the strong scintillation regime and we have utilized the scintillation modulation index to distinguish between the two classes of FRB models.

We provide in the next section some estimates for source size, scintillation bandwidth and the flux variation amplitudes. The scintillation bandwidth is inversely related to the scatter-broadening time $\delta \nu_{\rm sc}\approx (2\pi \delta t_s)^{-1}$ (see e.g. \citealt{1971MNRAS.155...51S, 1974ApJ...190..667B,Narayan1992}). In the limit of large flux modulations (i.e. order unity), this is the same as the observationally defined coherence bandwidth. However, as discussed in detail in this work, for an extended source, the flux modulations are smaller and the scintillation bandwidth is the inverse of the scatter-broadening time. In \S3, we discuss various factors that could affect the scintillation modulation index, such as two distinct scattering screens in the FRB host galaxy (one of which is near the source and the other being the host galaxy ISM). A brief discussion of FRB scintillation data is provided in \S4, and the main results of this work are summarized in \S5.

\section{Source size and flux variation beyond scintillation bandwidth}

We describe in \S\ref{sec:main-idea} how we can use flux variation over a frequency interval greater than the scintillation bandwidth to determine the size of the source when the plasma screen causing the scintillation is in the FRB host galaxy.
In \S\ref{section:source-size} we discuss the expected source size for two different FRB radiation models: magnetospheric origin and the faraway class of models.

\subsection{FRB source size from scintillation}
\label{sec:main-idea}
\begin{figure*}
\includegraphics[width = 0.9\textwidth]{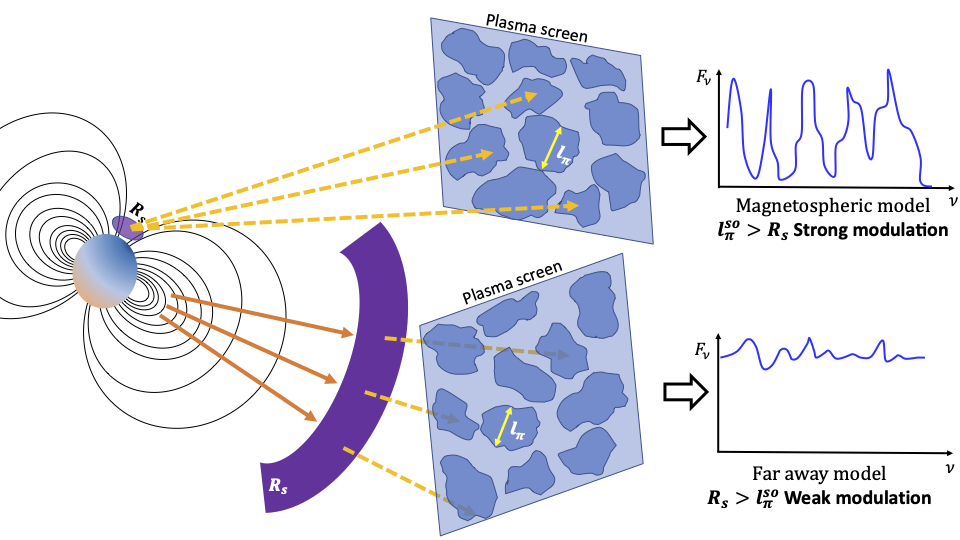}

\caption{Schematic sketch (not to scale) of an FRB source, whose radiation is scattered by a plasma screen in its host galaxy. Top - a magnetospheric model - the diffractive scale of the scattering screen (as projected on the source), $\ell_{\pi}^{\rm so}$, is large compared to the transverse size of the source, $R_{\rm s}$. In this case a strong modulation of the flux with frequency is observed due to scintillation. Bottom - for far away models for FRB radiation, $\ell_{\pi}^{\rm so}<R_{\rm s}$. In this case scintillation is suppressed due to the finite size of the source. }
    \label{fig:schem}
\end{figure*}

An intrinsically steady astronomical source can appear to fluctuate in brightness over time due to scintillation. This happens when the scattering plasma screen moves in relation to the source and observer by a distance $\ell_\pi$, the diffractive scale for the screen\footnote{The diffractive scale ($\ell_\pi$) is the transverse length in the plasma screen such that a wave suffers a differential phase shift of $\sim \pi$ rad across $\ell_\pi$ after 
crossing the screen. It should be noted that instead of the phase shift of $\pi$ that we have adopted in this work, the diffractive scale is often defined as a phase shift of one.}, projected on the source plane, causing the flux from the object to twinkle like stars. Short-duration bursts, like FRBs, don't twinkle because the time for the screen to move by $\ell_\pi$ is longer than their duration. However, sustained scintillations due to a scattering screen in the MW have been observed for the highly active FRB 20201124A \citep{Main2022}.
Moreover, scattering also affects short duration {electromagnetic (EM) bursts by spreading them out in time, due to propagation of the signal via multiple paths, and introducing stochastic fluctuations in their spectrum. Multi-path propagation introduces delays that manifest as temporal broadening for some FRBs, with a duration of approximately $\delta t_s \simeq 0.1$ to $> 1$ ms. However, other FRBs exhibit no broadening because it is too small to be measured as a broadening of the emitted burst width. The effect of multi-path propagation in these cases (in the strong scattering regime) would manifest as frequency structure in the spectrum and should be identifiable in the data with a frequency scale of $\delta\nu \sim (2\pi \delta t_s)^{-1}$. So far, pulse broadening for most FRBs where it is measurable is from the host galaxy whereas frequency structure has been identified in only a few cases, and it is from Milky Way scattering, e.g. \citealt{CHIME2018,Cordes2022}. Identifying the frequency structure and measuring the modulation index of the auto-correlation function due to scintillation in the host galaxy can help distinguish between different models of FRB radiation. 

An astronomical source is strongly scintillated with flux modulation amplitude of order unity when the diffractive scale for the screen, $\ell_\pi$, projected on the source plane, i.e. $\ell_\pi^{\rm so} = \ell_\pi d_{\rm SO}/d_{\rm LO}$, is larger
than the transverse size of the source ($R_s$), and $\ell_\pi$ is smaller than the Fresnel scale $R_F$ defined as
\begin{equation}
    R_F \equiv \left[{\lambda d_{\rm LO} d_{\rm SL}\over d_{\rm SO}}\right]^{1/2}, 
    \label{Rf}
\end{equation}
where $d_{\rm SO}$ is the distance between the source and the observer, $d_{\rm LO}$ is the distance between the plasma screen and the observer, and $d_{\rm SL}$ is distance between the source and plasma screen.

The scintillation bandwidth of the stochastic spectral fluctuation is inversely proportional to the
pulse broadening 
time of the scattering screen. The scintillation coherence bandwidth ($\delta\nu_{\rm sc}$) is typically obtained by fitting the auto-correlation function (ACF) of the FRB spectrum with a Lorentzian profile, which applies to a thin screen with a square-law structure function (e.g. \citealt{Cordes1998,Lorimer2004,Masui+15}) 
\begin{equation}
    r(\delta\nu) = {m_I^2 \over 1 + (\delta\nu/\delta\nu_{\rm sc})^2},
\end{equation}
where $m_I$ is the modulation or scintillation index which is the contrast between the flux of bright and dark scintles whose value is between 0 and 1, and the ACF is defined as
\begin{equation}
 r(\delta \nu)=\frac{1}{2\Delta \nu}\int_{\nu_1-\Delta\nu}^{\nu_1 + \Delta\nu} d\nu \left[\frac{f(\nu)}{\bar{f}(\nu)}-1\right]\left[\frac{f(\nu+\delta \nu)}{\bar{f}(\nu+\delta \nu)}-1\right]
\end{equation}
with $f(\nu)$ the receiver frequency response corrected flux at $\nu$ and $\bar f(\nu)$ is the corrected flux at $\nu$ averaged over the burst duration; it is assumed that any background flux has been subtracted from those values\footnote{In practice, filtering out noise and the calculation of auto-correlation function is more involved, and the main steps are as follows. Let $f(\nu, t)$ be the measured dynamic spectrum that contains a burst and a noise contribution.  Both are affected by the receiver frequency response $b(\nu)$. The off-burst data are used to estimate $f_{\rm off}(\nu) = N_0 b(\nu)$ where $N_0$ is the noise level that can be estimated from the off-burst data. Usually there is much more off-burst data to use so that noise fluctuations are averaged down and $b(\nu)$ is determined accurately. Then the on-burst dynamic spectrum is estimated as $f_{\rm on}(\nu, t) =  [f(\nu, t) - a b(\nu)] / b(\nu)$ where $a$ is a suitable constant so that the shape $b(\nu)$ matches the noise part of the on-burst data.  Next, $f_{\rm on}$ is summed over the time range containing the burst to  give the on-burst spectrum, $f_{\rm on}(\nu)$.  The on-burst spectrum is autocorrelated to finally yield the ACF.}. 

The modulation index $m_I$ depends on the relative sizes of $R_s$ and $\ell_\pi$ as well as whether the source is coherent across its visible size. For an incoherent source, or a source that consists of many patches of coherent regions unrelated to each other, the scintillation amplitude $m_I\sim \min\{\ell_\pi/R_s,1\}$. Some of these results are well-known (e.g. \citealt{Narayan1992}), however, for completeness, we provide the derivations of the results needed for this work in appendix \S \ref{sec:extendedsources}.

Thus, determining the scintillation index $m_I$ will enable us to measure the size of the FRB source and distinguish between the two broad classes of FRB models. This is shown schematically in Fig. \ref{fig:schem}.

We suggest selecting those FRBs that have scatter-broadened pulses dominated by scattering in the host galaxy and not the Milky Way's ISM.
The source size can be constrained by measuring the scintillation index, $m_I$, of the spectrum. If this index is smaller than unity, and the scintillation is not in the weak regime (where even for a point source $m_I < 1$), that would suggest that the source size is larger than the projected $\ell_\pi$ on the source plane. By measuring the frequency at which the modulation index $m_I$ transitions from order unity to below one, the source size can be determined accurately. This information can be used to narrow down the class of plausible radiation mechanisms for FRBs. We note that $m_I\ll 1$ is also obtained in the weak scintillation limit, which could be the case if an FRB source is outside the ISM of the host galaxy, as in the case of FRB 20200120E that is localised to a globular cluster in M81 \citep{Bhardwaj2021, Kirsten2022}. However, for weak scatterings, the scatter broadening is of the order of the wave period and the intensity scintillations are wideband, 
$\delta\nu_{\rm sc} /\nu \sim 1$, and thus it can be distinguished from strong scintillation with a low $m_I$ case. The fact that the scintillation bandwidth is the inverse of the scatter broadening time also means that in the weak scintillation case the scintillation bandwidth is very large, of order the central frequency of observations, and that should enable observers to remove the ambiguity between the two cases.

The diffractive scale, $\ell_\pi$, for a Kolmogorov spectrum of turbulence, is (see e.g. \citealt{Luan2014, BK2020}, for a derivation)
\begin{equation}
\ell_{\pi} \sim \left( {m_{\rm e} c^2\over q^2 n_{\rm e}\lambda}\right)^{{6\over5}} {\ell_{\rm max}^{{2\over 5}} \over L^{{3\over 5}} } \sim (2{\rm x}10^{10} {\rm cm})\, n_{\rm e,-2}^{-{6\over5}} L_{21}^{-{1\over 5}} \nu_9^{6\over 5} \left( {\ell_{\rm max,18}\over L_{21}}\right)^{2\over 5},
   \label{lpi}
\end{equation}
where $L$ is the thickness of the plasma screen, and $\ell_{\rm max}$ is the outer scale of the turbulence or the size of largest eddies in the scattering screen, $q$ \& $m_{\rm e}$ are the electron charge and mass and $n_{\rm e}$ is the electron density. The diffractive scale can be expressed in terms of the observed scatter-broadening of FRB pulses at 1 GHz ($\delta t_s$) or equivalently the scintillation bandwidth at that frequency,
\begin{eqnarray}
\label{lpi2}
   &\ell_\pi \sim {c\over \nu} \sqrt{ {d_{\rm L}\over 2 c \delta t_s}} \sim
        (1.2{\rm x 10^{10}\, cm}) \, \nu_9^{1.2} [d_{\rm L,21}/\delta t_{s,-7}]^{1/2}=\\
        & \sim
        (9.7{\rm x 10^{9}\, cm}) \, \nu_9^{1.2} [d_{\rm L,21}\delta\nu_{\rm sc,6}]^{1/2} \nonumber,
\end{eqnarray}
which has the distinct advantage that this expression for $\ell_\pi$ depends weakly on only one unknown parameter namely $d_L \equiv \min\{ d_{\rm LO}, d_{\rm SL}\}$ (the minimum of the distance between the observer and the scattering screen and source and the screen), as opposed to equation (\ref{lpi}) that contains three unknown parameters. We note that the width of the plasma screen ($L$) is approximately equal to $d_{\rm L}$.

The projected diffractive scale for scintillation caused by the MW-ISM  for cosmological sources is $\ell_\pi^{\rm so} \sim \ell_\pi d_{\rm SO}/d_{\rm LO} \sim 10^{17}\nu^{6/5}$cm; $\delta t_s\sim 40$ ns \& $d_{\rm SO}/d_{\rm LO}\sim 10^7$. This large size means that FRB sources, for all proposed models, can be considered effectively point objects when it comes to MW-ISM scintillation. Therefore, the MW-ISM is not useful for studying the FRB mechanism as discussed in this work.

Equation (\ref{lpi}) shows that the scale for scintillation in the host galaxy of an FRB is $\ell_\pi^{\rm so} \sim \ell_\pi\sim 10^{10}$cm. This scale falls between the sizes of FRB sources for the two classes of models (see \S\ref{section:source-size} for a detailed discussion). Therefore, scintillation in the host galaxy is well suited for the study of the FRB radiation mechanism.

The diffraction scale $\ell_\pi$ can also be expressed in terms of the scattering measure, SM, which has long been used to express scattering and scintillation quantities for Galactic pulsars, as follows
\begin{eqnarray}
\ell_\pi = 3.2 \times 10^9\ {\rm cm}\, 
	\left[  
		\left(\frac{\nu}{1\ \rm GHz} \right)^2   
		\left(\frac{10^{-3.5} \ \rm kpc \ m^{-20/3}}{\rm SM} \right)  
	\right]^{3/5}
\end{eqnarray}
where the nominal value for SM in the above equation is typical for a pulsar that is about 1 kpc away from us. Scintillation time scales for Galactic pulsars are of the order of hundreds to thousands of seconds, which is consistent with the value of $\ell_\pi$ given in the above equation, together with the effective transverse velocities of pulsars with respect to us ranging from tens to hundreds of km/s.

\subsection{FRB source sizes for different classes of models and their associated scintillation amplitude variations} 
\label{section:source-size}

\subsubsection{FRB source size - magnetospheric models}
\label{sec:magnetosphere}
The possible range for FRB source sizes for magnetospheric models can be determined using two general considerations. One consideration is based on the burst's variability time, while the other considers the maximum transverse size set by causality and coherence.

For FRB radiation originating in the magnetosphere, the observed variability of the lightcurve is likely dictated by the time-dependent activity of the {\it central engine}. However, the limiting factor is the transverse size of the source, $R_s$, which sets an upper limit on how fast the flux can vary with time. Let us consider that the radiation is produced at a distance $d$ from the NS surface by plasma moving toward the observer with Lorentz factor $\gamma$. In this case, most photons are beamed along the velocity vector of the plasma in a cone of opening angle $2\gamma^{-1}$. Therefore, the surface area visible to the observer has a radius no larger than $d/\gamma$. Moreover, due to the averaging of flux across this visible area, the observed lightcurve cannot vary on a time scale smaller than $d/(2c\gamma^2)$, even if the central engine were to vary much faster. Thus, $R_s \lta (2 d c \delta t_{_{\rm FRB}})^{1/2}$; where $\delta t_{_{\rm FRB}}$ is the observed variability time of an FRB. Considering that many FRBs have a variability time, or time for the rise of the lightcurve, that is tens of $\mu$s \citep[e.g., ][]{Nimmo2021}, the transverse size of the source has to be $\lta 10^7 d_8^{1/2}$cm; this estimate includes the effect of relativistic motion.

The other consideration stems from the coherent nature of FRB radiation. Waves produced by two particles separated by more than the wavelength of the wave ($\lambda$), in the comoving frame of the plasma, cannot radiate in phase and produce coherent radiation. Therefore, the natural transverse size of a ``coherent patch of the source region'' in the observer frame is $l_p = \lambda\gamma$. It is possible that the source consists of many independent coherent patches. In that case, the visible size of the source would be larger than $l_p$ by some factor that is hard to predict with confidence and is dependent on the precise specifics of the magnetospheric model. Taking a lower bound on $\gamma$ given by the lightcurve variability considerations discussed above, i.e. $\gamma \gta (d/2c \delta t_{_{\rm FRB}})^{1/2}$, we find a lower limit on the source size $R_s \gta 10^3 \nu_9^{-1} d_8^{1/2}$cm. 

The estimate for the source size provided in the preceding paragraphs shows that the expected $R_s$ for magnetospheric models would be much smaller than $\ell_\pi$ in the ISM of the host galaxy. Additionally, the plasma within a parsec of the FRB source is unlikely to have an $\ell_\pi$ as small as 10$^7$ cm. This is because for a plasma screen at $\lta 10^{18}$ cm with $\ell_\pi \sim 10^7$ cm, the required electron density would be $\gta 0.4~$cm$^{-3}$ (according to equation \ref{lpi}). This required electron density is much larger than what is realistic for the region within approximately one parsec of the magnetar that is dominated by its wind. This is explained below and discussed in greater detail in \S\ref{sec:two-screen}.

Assuming a mass loss rate of $\dot M$ in the form of an $e^\pm$ relativistic wind with a Lorentz factor of $\gamma_w$, the comoving particle density at a distance $D_1$ from the central object is
\begin{equation}
    n_{\rm e}'(D_1) = (3{\rm x10^{-5}\, cm^{-3}})\, {\dot M}_{10} D_{1,15}^{-2} \gamma_w^{-1},
    \label{wind-density}
\end{equation}
where ${\dot M}_{10}$ is the mass loss rate in units of $10^{10} {\rm g\; s^{-1}}$.
The required density of $0.4$ cm$^{-3}$ at a distance of $10^{18}$ cm corresponds to a mass loss rate of approximately $10^{20}$ g s$^{-1}$. This mass loss rate is of the same order as the winds observed in O-stars (e.g. \citealt{Puls2008}) and several orders of magnitude larger than what would be expected from magnetars. Therefore, it is unlikely that a screen exists within $\sim 10^{-2}$pc of the NS with an $\ell_\pi$ of less than $10^7$cm if the medium is dominated by the magnetar wind, which it almost certainly is for a NS capable of producing an FRB.

However, the density can be this large or larger for a supernova remnant at a distance between $\sim 0.1$ pc and a few pc from the magnetar. For instance, the Crab Nebula has dense filaments with densities $\gtrsim 10^4$~cm$^{-3}$ and is known to significantly scatter radio waves from the pulsar with a great deal of variability due to motions of the filaments \citep[e.g.,][]{Driessen2019}. Due to the lack of information regarding the age of the system, there is significant uncertainty in determining the distance, density, and ionization fraction of the supernova remnant surrounding the FRB source.
We discuss in \S\ref{sec:two-screen} the constraints on a supernova scattering screen from considerations such as the free-free optical depth to radio photons and the maximum contribution to the dispersion measure (DM) from the FRB host galaxy. 

In summary, for the case of FRBs originating in the magnetosphere, the amplitude of flux variation across frequency scales greater than the scintillation bandwidth for any realistic scattering screen is likely to be of the order of the mean flux.

 \subsubsection{FRB source size - far away models}
For the far away model, the duration of the burst is determined by the time it takes for the shock front's Lorentz factor to decrease by a factor $\sim 2$, since the frequency and the luminosity of the emergent coherent radiation depend on this Lorentz factor. Using arguments similar to the magnetospheric case, we can derive an expression for the observed FRB duration, ignoring the redshift factor: $t_{\rm FRB} \sim d/(2 c \gamma^2)$, where $d$ is the distance to the source and $\gamma$ is the Lorentz factor of the shock \footnote{ The difference from the limit in the magnetospheric case is that here we have the overall FRB duration rather than its variability time-scale. That being said, in far-away models, one typically expects $\delta t_{\rm FRB}\approx t_{\rm FRB}$ \citep{BK2020}, with departures from this approximated equality coming at the cost of significant reduction in the observed radiative efficiency.}. Therefore, $\gamma \sim (d/2c t_{\rm FRB})^{1/2}$, and the transverse source size is
\begin{equation}
\label{eq:Rs}
R_s = \frac{d}{\gamma} \approx 2c\, t_{\rm FRB} \, \gamma \approx \sqrt{2c t_{\rm FRB} d},
\end{equation}
or
\begin{equation}
    R_s \approx (7.7\times 10^{9} \mbox{ cm})\, t_{\rm FRB,-3}^{1/2}d_{12}^{1/2} \approx (6\times 10^{9} \mbox{ cm})\, t_{\rm FRB,-3}\,\gamma_{_2}
\end{equation}
The scintillation amplitude
falls below unity when $R_s \gtrsim \ell_\pi^{\rm so}$. That means that if the FRB radiation is generated at a distance further than a critical distance,
\begin{equation}
  d_{\rm crit} = (4\times10^{11} {\rm cm})\, t_{\rm FRB,-3}^{-1} n_{\rm e,-2}^{-{12\over5}} \nu_9^{12\over 5} L_{22}^{-2/5} \left( \frac{\ell_{\rm max,18}}{L_{22}} \right)^{4/5}
  \label{eq:dlim}
\end{equation}
then the flux variation amplitude over frequency scales greater than the scintillation bandwidth due to scattering by plasma in the host galaxy is suppressed below unity. This is something that observers can look for in the data to provide a direct constraint on the otherwise highly uncertain FRB radiation mechanism.

We define a characteristic frequency, $\nu_{\rm so}$, below which the scintillation amplitude is suppressed by the finite source-size of the far-away model of FRBs. Using Eq. \ref{eq:dlim} we obtain
\begin{equation}
   \label{eq:nusou}
    \nu_{\rm so} = \mbox{(1.4 GHz)}\; t_{\rm FRB,-3}^{5/12} n_{\rm e,-2} d_{12}^{5/12} L_{22}^{1/6} \left(\frac{\ell_{\rm max,18}}{L_{22}}\right)^{-1/3}
\end{equation}

All the cases of scatterings mentioned above lie in the so-called strong scattering regime,
 i.e. $\ell_{\pi} < R_F$.
The strong scattering happens below the following critical frequency
\begin{equation}
	\label{eq:nustar}
    \nu_* = \mbox{ (15.7 GHz)}\;  n_{\rm e,-2}^{12/17} L_{22}^{7/17} 
    \left(\frac{\ell_{\rm max,18}}{L_{22}}\right)^{-4/17} .
\end{equation}
where, $R_F = \sqrt{\lambda d_{\rm SL}}$ from Eq. \ref{Rf}, as $d_{\rm LO} \approx d_{\rm SO}$ for the scattering screen located in the host galaxy, and $L\approx d_{\rm SL}$. If the transverse source size $R_s$ is less than $R_F$, then $\nu_{\rm so} < \nu_*$. 

The scattering time, or the amount by which an FRB pulse is broadened, is
\begin{equation}
    \delta t_s =  \frac{R_{\rm F}^2 (\delta \theta)^2}{2c\lambda} = (1.9\times10^{-6} s) \nu_9^{-22/5} n_{\rm e,-2}^{12/5} L_{22}^{3/5} \ell_{\rm max,18}^{-4/5}
    \label{eq:tscat}
\end{equation}

Since $\delta t_s$ is strongly dependent on the observed frequency, it is useful to introduce a new variable that removes this dependence on $\nu$: $\delta t_{0} = \delta t_s \nu_{9}^{22/5}$. If $\nu_{\rm so}$, $\nu_*$, and $\delta t_s$ could be measured for a burst, then with the relations given by Eqns. \ref{eq:nusou}, \ref{eq:nustar} and \ref{eq:tscat}, we can obtain the following parameters for the FRB source and the scattering screen when coherent radio waves are generated outside the magnetosphere:
\begin{equation}
    L = (5.6 \times 10 ^{21} \mbox{ cm}) \delta t_{0,-6}^{-5/8} \nu_{*,10}^{17/8}
    \label{eq:Lcalc}
\end{equation}
\begin{equation}
    d = (1.15\times10^{12} \mbox{ cm}) t_{\rm FRB,-3}^{-1} \delta t_{0,-6}^{-5/8} \nu_{\rm so,9}^{12/5} \nu_{*,10}^{-51/40}
    \label{eq:dcalc}
\end{equation}
\begin{equation}
    n_{\rm e} = (7.4 \times 10^{-3} \mbox{ cm}^{-3}) \nu_{*,10}^{17/96} \delta t_{0,-6}^{35/96} \left( \frac{\ell_{\rm max,18}}{L_{22}} \right)^{1/3}
    \label{eq:ncalc}
\end{equation}
We note that the re-scaled scattering time, $\delta t_0$ is independent of the observed frequency and hence Eqns. \ref{eq:Lcalc}, \ref{eq:dcalc} and \ref{eq:ncalc} don't have any dependence on the frequency, as  expected. In Fig. \ref{fig:quants}, we plot how $L$, $d$, and $n_{\rm e}$ vary with some observational parameters.

There are three different scenarios possible for the far-away model for FRBs. Depending on the frequency band for the observation, we may observe either strong scintillation with weakly modulated flux amplitude (for $\nu < \nu_{\rm so} < \nu_*$), strong scintillation with flux fluctuation of the order of the baseline flux ($\nu_{\rm so} < \nu <\nu_*$), or weak scintillation with $\delta f/f < 1$ ($\nu > \nu_*$). The factors that contribute to $\delta f/f$ are summarized in \ref{sec:fluxfluctuations}.

In summary, if it is found that the ACF amplitude at some frequency is of order unity, then that would narrow down the possible FRB mechanism to a magnetospheric origin. On the other hand, if the amplitude turns out to be less than unity and the scintillation bandwidth is resolved by the detector, then that would support the far-away model (as long as $\nu < \nu_*$). In the latter case, the determination of $\nu_{\rm so}$ and $\nu_*$, together with the scattering width of pulses, could be used to determine the distance from the NS where FRB radiation is produced (for the far-away model for FRBs), as well as the density and distance of the scattering screen in the host galaxy. 

\begin{figure*}
\includegraphics[width = \textwidth]{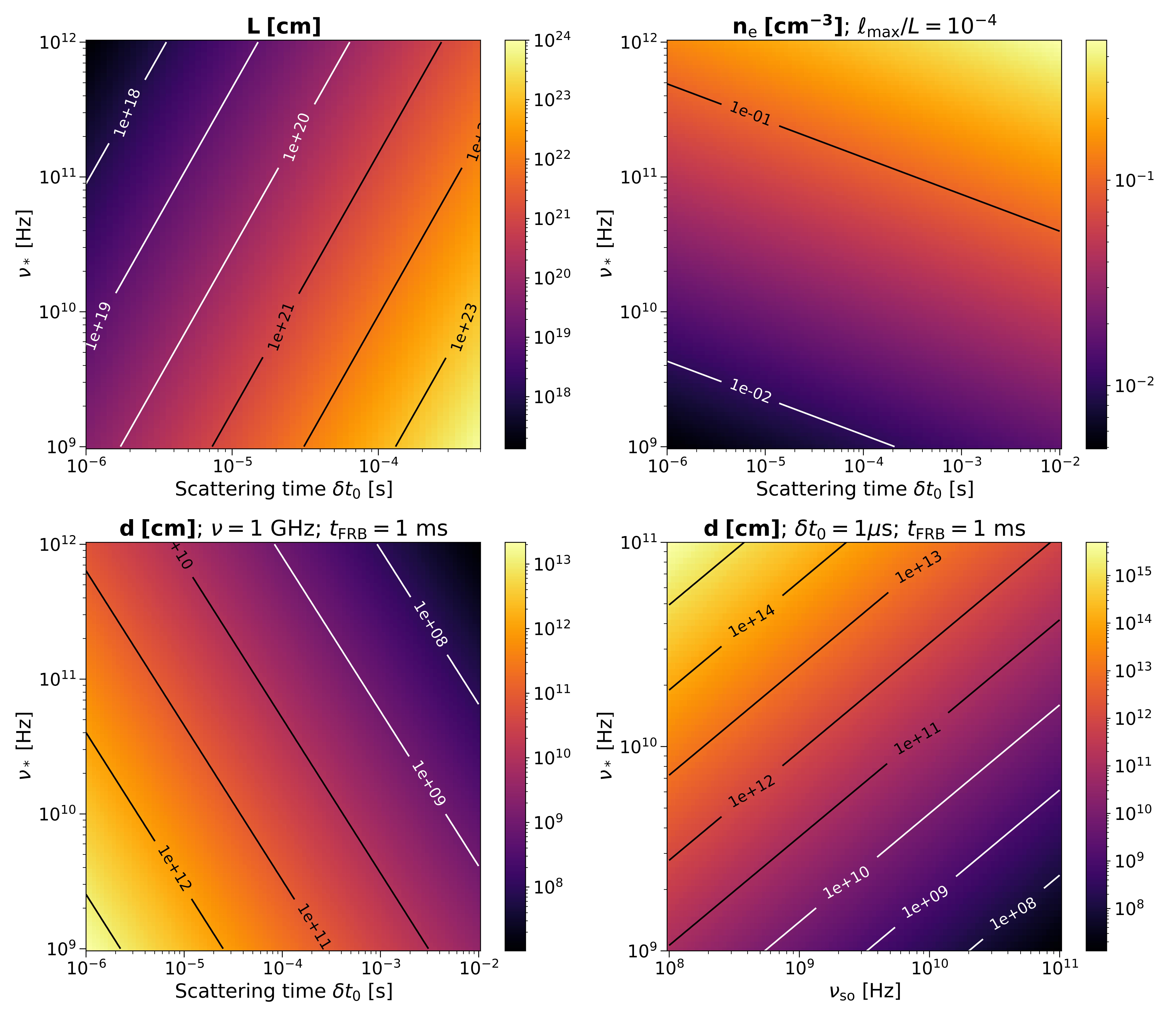}
    \caption{Plots of (a) the thickness of the scattering screen L ($L\sim  d_{\rm SL}$), (b) the electron density in the scattering screen $n_{\rm e}$, (c) the distance of the emitting region from the central engine $d$, as a function of the re-scaled scattering time $\delta t_0$ and $\nu_*$, and (d) $d$ as a function of $\nu_{\rm so}$ and $\nu_*$. The respective scales for these quantities are given both by the colour bars on the right side of each panel and the plot contours depicting order-of-magnitude changes in colour.}
    \label{fig:quants}
\end{figure*} 

Besides scintillation, the fragmentation instability of an FRB pulse can also cause variability in its light-curve and limit the coherence bandwidth of its spectrum. This instability arises due to the interaction of a high amplitude EM pulse with electrons in the medium through which it propagates \citep{Max1974}. It leads to the fragmentation of the pulse into honeycomb cells, and we explore its growth and impact on coherence bandwidth \& ACF amplitude in \S\ref{fragmentation-instability}.

\subsection{Fragmentation instability of FRB pulse, and its effects on the coherence bandwidth}
\label{fragmentation-instability}

The coherent FRB radiation pulse fragments in the longitudinal and transverse directions due to the development of an instability that results from the interaction with the plasma in its path. This instability causes the FRB pulse to break up into honeycomb cells when it reaches a nonlinear stage. The segments of the fragmented pulse spread laterally (diffractive spreading), and if the angular size of the cells at their formation site is smaller than the diffractive scale, the observer will receive signals from multiple fragments. The interference of these signals will imprint a structure in the FRB spectrum. We calculate the coherence bandwidth due to the fragmentation instability and assess whether this might pose difficulties in identifying diffractive scintillation signals in the FRB host galaxy.

We make use of the results presented in \citet{sobacchi21_modulational_instability} who have analyzed the fragmentation instability as the EM wave moves through an electron-ion plasma. The transverse and longitudinal wave-numbers $(k_\perp, k_\parallel)$ of the fastest growing modes of this instability are
\begin{equation}
    c k_\perp \approx a_0 \omega_p,  \quad  c k_\parallel \approx \min\left\{ a_0\omega, \omega_p\right\},
    \label{unstable-k}
\end{equation}
where $\omega_p=(4\pi q^2 n_{\rm e}/m_{\rm e})^{1/2}$ is electron plasma frequency, $\omega$ is FRB wave frequency, 
\begin{equation}
    a_0 = {qE_0\over m_{\rm e}c\omega}
    \label{a0}
\end{equation}
is a dimensionless strength parameter for the FRB pulse, $E_0=\sqrt{L_{\rm FRB}/cR^2}$ is the electric field strength associated with the radio pulse, $L_{\rm FRB}$ is FRB luminosity, and $R$ is the distance from the central object where the plasma responsible for the instability resides. The growth rate for the instability is
\begin{equation}
    \Gamma \approx {a_0^2\omega_p^2\over \omega}.
    \label{growth-rate}
\end{equation}

The angular size of fragments (asymptotically), after their lateral spreading, is 
\begin{equation}
    \theta_d = {c k_\perp\over \omega} \sim \sqrt{ {\Gamma\over\omega}},
\end{equation}
where we made use of equations \ref{unstable-k} \& \ref{growth-rate}. A faraway observer would receive photons from many different segments as long as $\theta_d > 2\pi/(k_\perp R)$, and thus a very narrow pulse would be temporally broadened to the duration
\begin{equation}
    t_d \approx {R \theta_d^2\over 2 c} \sim {R\Gamma\over 2c\omega}
\end{equation}
due to the difference in arrival time of photons that have followed different geometrical paths. Since $\Gamma R/c$ is the e-folding factor of the instability as it operates over the time $R/c$, we see that the pulse broadening time is the wave period times this growth factor.

The interference of waves that arrive at the observer having traveled along different paths causes fluctuation of the observed spectrum. The frequency bandwidth for this fluctuation is given by the Fourier theorem, i.e.
\begin{equation}
    \Delta\nu_{nfi} \approx {1\over 2\pi t_d} \sim {\nu\over (R\Gamma/c)}.
\end{equation}
We see from the above equation that the coherence bandwidth due to the instability is $\sim 0.1$ GHz at $\nu=1$ GHz when the instability growth factor is $\sim 10$.

The transverse size of the fragmented FRB pulse, using the particle density in the wind (eq. \ref{wind-density} modified to consider electron-proton plasma), is
\begin{equation}
    \ell_\perp \sim {2\pi\over k_\perp} = {2\pi c\over a_0 \omega_p} \sim (1.6{\rm x10^8 cm}) {R_{13}^2 \gamma_w^{1/2} \nu_9\over L_{\rm FRB,42}^{1/2} {\dot M}_{10}^{1/2} }.
    \label{l-perp2}
\end{equation}
And the angular size of fragments increases due to wave diffraction to the following asymptotic value
\begin{equation}
    \theta_d \sim (2{\rm x10^{-7} rad}) \; {L_{\rm FRB,42}^{1/2} {\dot M}_{10}^{1/2} \over R_{13}^2 \gamma_w^{1/2} \nu_9^2}.
\end{equation}
The FRB pulse is scatter-broadened temporally to 
\begin{equation}
    t_d \sim {R \theta_d^2\over 2c} \sim {\eta \over 2\omega} \gta \nu^{-1},
\end{equation}
where $\eta$ is the e-folding factor by which the instability grows at radius $R$
\begin{equation}
    \eta \equiv {\Gamma' R\over c\gamma_w} \sim 0.1 {L_{\rm FRB,42} \dot M_{10}\over \nu_9^3 R_{13}^3 \gamma_w}.
    \label{eta2}
\end{equation}
The fragmentation instability results presented here apply for the case where $a_0 \ll 1$, and therefore are only valid for $R\gta 10^{13}$cm.

The coherence bandwidth due to interference between photons from different segments of the broken-up FRB pulse is
\begin{equation}
    \delta\nu_d \sim {1\over 2\pi t_d} \sim {2\nu\over \eta}.
    \label{nu-coh-inst}
\end{equation}
The frequency dependence of the coherence bandwidth is $\omega^{-4} L_{\rm FRB}(\nu)$. Depending on the FRB spectral luminosity, $L_{\rm FRB}(\nu)$, this might be similar to the spectral dependence in strong scintillation. However, for a given FRB, $L_{\rm FRB}(\nu)$ is observable, and the resulting spectral dependence may be distinguishable from diffractive scintillation.

The amplitude of flux variation across the coherence band of $\delta\nu_d$ depends on the relative sizes of the FRB source and the FRB pulse fragments. For a source of finite size, the amplitude of flux fluctuation across $\delta\nu_d$ is reduced for the same reason as in the case for scintillation --  two points in the source separated by a distance larger than $\ell_\perp$ suffer a phase shift of approximately $\pi$ radians due to the slightly different paths taken by the wave from the source to the observer.  This leads to the superposition of fluctuating fluxes from different patches of the source of size $\sim\ell_\perp$, resulting in a dimensionless flux variation amplitude across $\delta\nu_d$ of 
\begin{equation}
 m_{I} = \delta f/f \sim \min\left\{ 1, \ell_\perp/R_s \right\}.
\end{equation}

From equations \ref{unstable-k} and \ref{growth-rate}, it can be seen that $\ell_\perp \sim (2\pi\lambda R)^{1/2}/\eta^{1/2} = \sqrt{2\pi}R_F/\eta^{1/2}$. Thus, a fully developed fragmentation instability corresponds to the strong scintillation case, and they have the same effect on coherence bandwidth. When $\eta \gg 10$, the fragments are much smaller than the Fresnel scale, and as a result, the interference of waves from multiple fragments is observed.  This means that these fragments serve the same purpose as the diffractive scale in limiting the source size under conditions of strong scintillation. We see from equation (\ref{eta2}) that for the fiducial parameters of the magnetar wind, and the luminosity of a typical non-repeating FRB, the instability can grow to non-linear stage and fragment the FRB pulse only at $R< 10^{13}$cm. The transverse size of the fragments from equation (\ref{l-perp2}) is found to be $\sim 10^{7}$cm for these parameters, and the coherence bandwidth of the pulse due to the instability (eq. \ref{nu-coh-inst}) is $\sim 0.1\, \nu_9$ GHz. Therefore, the wave diffraction resulting from these fragments is good for testing the validity of far away models where the source size is $>10^{9}$cm. An $e^\pm$ wind from the magnetar leads to stronger fragmentation instability of the FRB pulse, but that does not change the main conclusions described in this sub-section. 

\section{Factors that affect the scintillation modulation index}
\label{sec:effects-on-ACF}

A number of factors could affect the scintillation amplitude, which could in turn affect the determination of the FRB source size. One factor has already been mentioned in \S\ref{section:source-size}, which is the finite frequency resolution of the detector. Secondly, an FRB pulse traveling through the medium in the vicinity of the magnetar could undergo scattering and fragmentation, which can also affect the observed scintillation modulation index, $m_I$. For instance, the presence of a scattering screen within a few parsecs of the source, in addition to the ISM of the host galaxy that lies at an effective distance of $10^3$ pc, could modify $m_I$ compared to the single scattering screen model. This is discussed in \S\ref{sec:two-screen}. Finally, the effect of scintillation in the MW-ISM on the fluctuation amplitude $m_I$ is described in \S\ref{sec:MW-scintillation}. In \S \ref{sec:fluxfluctuations}, we summarize the different effects that can affect $m_I$ and provide a general expression for $\delta f/f$, incorporating the various considerations discussed in this work.

\subsection{Two plasma screens: one close to the source and another further away in the host galaxy}
\label{sec:two-screen}

In this subsection, we examine the possibility of a circum-stellar medium located closer to the source (at a distance of order 1 pc or less) that could increase the apparent size of the source, affecting scintillation by a screen at larger distances in the host galaxy. Observations point to the existence of such a screen, as evidenced by the presence of an extremely magneto-ionic environment around FRB 20121102 \citep{Michilli+18}. Additionally, there has been the detection of persistent radio sources with a non-thermal origin at the localized positions of certain FRBs \citep{Chatterjee+17, Marcote2017, Niu2022}. This close-in screen will imprint its own scintillation bandwidth on the FRB spectrum, depending on the source size and the diffractive length for this screen. We estimate how this screen affects the limits that observations can place on the source size.

Let us consider a close-in scattering screen at a distance $D_1$ from the source (we use the index 1 to refer to the closer screen); $D_1$ might be of order 10$^{18}$cm or less. We note that the standard scintillation theory needs to be modified due to the non-linear wave amplitude of the FRB radiation at short distances from the source -- the nonlinearity is quantified by a dimensionless parameter $a_0$ (eq. \ref{a0}) --  but, for simplicity, we will disregard that effect. Our goal is to provide an order of magnitude estimate for the impact of the close-in screen. We take the diffractive scale for this screen to be $\ell_\pi^{(1)}$. Thus, the scintillation bandwidth for the screen is: 
\begin{equation}
    \delta\nu^{(1)}_{\rm sc} \approx \left[ { \min\left\{\ell_\pi^{(1)}, R_F^{(1)}\right\}\over \lambda} \right]^2 {c\over \pi D_1}, \label{eq:dnusc1}
\end{equation}
where 
\begin{equation}
    R_F^{(1)} = \sqrt{D_1 \lambda}
\end{equation}
is the Fresnel radius for the close-in scattering screen.

A point source is broadened to size
\begin{equation}
 R_s^{(1)} \sim D_1 \left[ {\lambda\over \ell_\pi^{(1)} }\right] \label{eq:rs1}
\end{equation}
by scintillation in this screen when $\ell_\pi^{(1)} < R_F$, i.e. in the strong scintillation regime. If the broadened source size were larger than the diffractive scale for the further-away screen ($\sim$kpc away), the close-in screen could hinder the ability of the further-away screen to resolve the source. The closer screen will imprint its own scintillation bandwidth ($\delta\nu^{(1)}_{\rm sc}$) on the observed spectrum.

Let us consider that $R_s^{(1)} = \xi \ell_\pi^{(2)}$ with $\xi > 1$; $\ell_\pi^{(2)}$ is the diffractive scale for the ISM scattering screen in the host galaxy that lies further away from the source than the screen at $D_1$. In this case, the closer screen magnifies the source length enough to suppress the scintillation modulation index caused by the further away screen.
The scintillation bandwidth for the close-in screen can be rewritten in terms of $\ell_\pi^{(2)}$ as
\begin{equation}
 \delta\nu^{(1)}_{\rm sc} \sim {\nu\over\pi} \left[ { R_F^{(1)} \over \xi\ell_\pi^{(2)}}\right]^2 .
\end{equation}
The scintillation bandwidth, $\delta\nu^{(1)}_{\rm sc}$, is capped at the wave frequency $\sim\nu$ when the scattering regime transitions from strong to weak. 
The scintillation bandwidth of the closer screen may not be observationally detectable when it is smaller than the frequency resolution of the radio receiver ($\delta\nu_{sc}^{(1)} < \delta\nu_{ob}$) as the signal would be wiped out by averaging over $\delta\nu_{ob}$. The signal is also not detectable if $\delta\nu^{(1)}_{sc}$ is of the order of or larger than the scintillation bandwidth for MW-ISM scintillation at high latitude ($\delta\nu_{sc}^{(1)} > \delta\nu_{G}\sim10^6$Hz). In such cases, it could be mistaken for MW-ISM scattering or intergalactic medium (IGM) scintillation.

This implies that, for a certain range of distances, the close-in screen can significantly amplify the source size, making it impossible to use scintillation in the ISM of the FRB host galaxy to differentiate between different classes of FRB models. Additionally, the impact of the close-in screen on the FRB spectrum could go unnoticed. When the plasma screen is closer to the source than $D_1^{\rm ob}$, the scintillation coherence bandwidth is smaller than $\delta\nu_{\rm ob}$, rendering it unobservable:
\begin{equation}
\begin{split}
    & \delta \nu_{sc}^{(1)} < \delta\nu_{\rm ob} \\
    & \implies D_1\lesssim D_1^{\rm ob} = {\pi \delta\nu_{\rm ob} \xi^2 \big[\ell_\pi^{(2)}\big]^2\over \nu\lambda} \sim 10^{10} {\rm cm}\,\left( {\delta \nu_{\rm ob,3}\over\nu_9}\right) {\xi^2 D_{2,21} \over \nu_9 \delta t_{s,-4}^{(2)} }
\end{split}
\end{equation}
where $D_2$ is the distance of the ISM scattering screen in the FRB host galaxy from the radio source, and $\delta t_{s}^{(2)}$ is the scatter broadening time for this screen. When the plasma screen is located at a distance greater than $D_1^G$, its scintillation coherence bandwidth becomes larger than $\delta\nu_G$, leading to confusion with the MW-ISM scattering screen or the IGM scattering. The distance $D_1^G$ is given by
\begin{equation}
    \delta \nu_{sc}^{(1)} > \delta\nu_G \implies D_1 \gta D_1^{G} = 10^{13} {\rm cm}\,\left( {\delta \nu_{G,6}\over\nu_9}\right) {\xi^2 D_{2,21} \over \nu_9 \delta t_{s,-4}^{(2)} }.
\end{equation}
A plasma screen located between $D_1^{\rm ob}$ and $D_1^G$ would broaden the source size to the extent that it becomes larger than $\ell_\pi^{(2)}$ for the host galaxy's ISM, thereby interfering with the ability of the further away screen to differentiate between the two classes of FRB models. However, it will imprint its own observable scintillation pattern on the FRB spectrum.  This scintillation pattern can be utilized to distinguish between the near-field and far-away FRB models in most cases. We demonstrate below that any screen situated at distances smaller than $D_1^{\rm ob}$, and capable of affecting the host galaxy's ISM screen, would be opaque to induced-Compton (IC) scattering. Hence, no screen with $D_1 < D_1^{\rm ob}$ presents any problem. Furthermore, we show that any realistic plasma screen located beyond $D_1^G$ can be independently used to differentiate between the various classes of FRB models.

For the scintillation by the close-in screen to be in the strong-scattering regime, the minimum electron density required can be determined using equation (\ref{lpi}),
\begin{equation}
    \ell_\pi^{(1)} < R_F^{(1)} \implies 
    n_{\rm e} > (50\, {\rm cm^{-3}})\, \nu_9^{17\over 12} D_{1,15}^{-{7\over12}} \left[ {\ell_{\rm max}^{(1)}\over D_1} \right]^{1/3},
    \label{ne-lim1}
\end{equation}
where $\ell_{\rm max}^{(1)}$ is the maximum eddy size for the close-in scattering screen.

Moreover, the condition $R_s^{(1)} =\xi \ell_\pi^{(2)}$ with $\xi>1$, places a limit on the diffractive scale for the close-in screen, which is
\begin{equation}
  \ell_\pi^{(1)} \sim (3{\rm x10^6\,cm})\, \nu_9^{-1} \xi^{-1} D_{1,15}/\ell_{\pi,10}^{(2)}.
  \label{l-pi1-lim}
\end{equation}
The electron density required for this diffractive scale can be obtained using equation (\ref{lpi}), and is given by
\begin{equation}
    n_{\rm e} \sim ({\rm 1.5x10^3\, cm^{-3})}\, 
     \xi^{5/6}\,\Big[\ell_{\pi,10}^{(2)}\Big]^{5/6}\,  D_{1,15}^{-1} \,\nu_9^{11/6} \left(\frac{\ell_{\rm max}^{(1)}}{D_1}\right)^{1/3}.
     \label{ne-lim2}
\end{equation} 
This requirement on electron density is more stringent than given by equation (\ref{ne-lim1}). By combining Eqs. \ref{ne-lim1} and \ref{ne-lim2}, we can obtain an upper limit on the distance between the first screen and the source that ensures that the first screen is both in the strong scintillation regime and capable of suppressing the scintillation caused by the second screen:
\begin{equation}
    D_1 \lta (3.5{\rm x}10^{18}\mbox{cm})\; \xi^2\, \nu_9 \left[\ell_{\pi,10}^{(2)}\right]^2.
    \label{D1-uplim}
\end{equation}

A lower limit on the distance between the first screen and the source can be placed based on the requirement that the screen should not be opaque to induced-Compton (IC) scatterings. The optical depth of the screen to IC scatterings when it has electron density $n_e$ is (e.g. \citealp{KumarLu-IC-scat-2020})
\begin{equation}
\label{eq-IC}
   \tau_{\rm IC} \approx {\sigma_T L_{\rm FRB} n_{\rm e} c t_{\rm FRB} \gamma_w \over 8\pi^2 D_1^2 m_{\rm e} \nu^3} \sim 4{\rm x}10^{-7} { L_{\rm FRB,42} t_{\rm FRB,-3} n_{\rm e} \gamma_w\over D_{1,15}^2 \nu_9^3},
\end{equation}
where $L_{\rm FRB}$ \& $t_{\rm FRB}$ are the FRB luminosity and duration, $\sigma_T$ is Thompson scattering cross-section, and $\gamma_w$ is the Lorentz factor of the medium through which the FRB pulse is moving at distance $D_1$ from the source. Given the requirement on electron density (equation \ref{ne-lim2}) for the close-in screen, we see that this screen would be IC opaque when its distance from the source is smaller than
\begin{equation}
    D_1^{\rm IC} \approx (10^{14}\mbox{cm})\; { L_{\rm FRB,42}^{1\over3} t_{\rm FRB,-3}^{1\over3} \xi^{5\over18} \left[\ell_{\pi,10}^{(2)}\right]^{5\over18}  \gamma_w^{1\over3} \over \nu_9^{7\over18}} \left[\frac{\ell_{\rm max}^{(1)}}{D_1^{\rm IC}}\right]^{1\over9}.
    \label{D1-lowlim}
\end{equation}
  
The bottom line is that the first screen must lie within the two distances provided by equations \ref{D1-uplim} and \ref{D1-lowlim} to allow the passage of FRB radiation, and substantially increase the angular size of the FRB source so that it hampers the second-screen's (host galaxy ISM scintillation) ability to distinguish between near-field and far-away models for FRBs. Placing the first screen at such an intermediate distance is difficult, as demonstrated below for two specific sources of plasma close to the neutron star.

Assuming the FRB source to be a magnetar, an additional constraint on the viability of the first screen is imposed by the mass loss rate from the magnetar, which affects the medium within a radius of approximately 0.1 pc. This constraint places a limit on the electron density. We show that the electron density in the wind of a typical magnetar is too low for the plasma screen within the radius of influence of the wind to significantly affect the scintillation caused by the further out screen in the host galaxy.

Furthermore, the column density of the first screen should not exceed the observational constraint on the host galaxy's contribution to the total DM. This constraint applies specifically to the case where the first screen is a remnant of the supernova explosion.

The medium within $\sim 10^{-1}$pc of the NS is likely to be the e$^\pm$ wind from it. The density requirement given by Eq. \ref{ne-lim2} then translates to the mass loss rate using equation (\ref{wind-density})
\begin{equation}
\label{eq:Mdottwoscreen}
    {\dot M} \sim (5{\rm x10^{17} g \, s^{-1}})\, \xi^{5/6}\,\Big[\ell_{\pi,10}^{(2)}\Big]^{5\over6} D_{1,15} \nu_9^{11\over6} \left(\frac{\ell_{\rm max}^{(1)}}{D_1}\right)^{1\over3}.
\end{equation}
This mass loss rate is of order a few percent of winds observed in O-stars (e.g. \citealt{Puls2008}) and several orders of magnitude larger than expected from even active magnetars. The constraint on the electron density imposed by the magnetar wind renders the possibility of the first-screen, located between the radius of approximately $10^{15}$ cm and $10^{18}$ cm and capable of interfering with the host galaxy ISM scintillation, unfeasible\footnote{The high electron density requirements can be fulfilled in cases where the wind is being supplied by a binary companion star. There is some evidence for this scenario, given the existence of a 16.3-day period for FRB 20180916B \citep{CHIME+20}. However, a very small fraction of FRBs show a periodic burst window that can be ascribed to a binary system. As it stands, such cases may be considered outliers in the observed FRB population.}.
 
There is, however, the possibility that at somewhat larger radius ($\gta10^{-1}$pc), the medium might not be the wind from the NS, and the electron density could be associated with the supernova remnant or flares from the magnetar accumulated over a long period of time. The results depend on the age of the supernova remnant (SNR), which determines the radius of the remnant, and the properties of the magnetar and its radiative and wind output histories that determine the degree of ionization of the remnant. A self-consistent calculation of all these effects together will need to be taken up in separate work. We assume here that the SNR age is more than a few tens of years so that its radius is $\gta10^{18}$cm, and the radiation from the magnetar and the shock interaction between the SNR and magnetar wind, as well as the ISM, have completely ionized the remnant. 

The electron density associated with a fully ionized SNR of mass ($m\, M_\odot$) that lies at distance $D_1$ from the source is
\begin{equation}
    n_{\rm e}^{\rm snr} \sim (10^5\, \rm cm^{-3})\, m\, D_{1,17}^{-3}.
\end{equation}
The contribution to the dispersion measure from the SNR is
\begin{equation}
    {\rm DM}_{\rm snr} \sim {n_{\rm e}^{\rm snr} D_1 \over 1 {\rm pc}} \sim
    (3{\rm x10^3\, pc\, cm^{-3}})\, m\, D_{1,17}^{-2},
\end{equation}
and the free-free absorption optical depth of the remnant is (\citealp{Rybicki79})
\begin{equation}
    \tau_{ff}(D_1) \approx \alpha_{ff} \, D_1 \approx 20\, m\,T_4^{-3/2} D_{1,17}^{-5} \nu_9^{-2}, 
\end{equation}
where $T_4$ is the temperature of the remnant in units of 10$^4$k. Thus, the SNR is optically thick to GHz waves, and its contribution to the FRB DM exceeds the limit on DM$_{\rm host}$ for most FRBs for $D_1 \lesssim 10^{18}$cm.

The limits imposed by the magnetar wind mass loss rate, as well as the optical depth and DM of the SNR, suggest that the close-in scattering screen, capable of broadening the FRB source size to the scale of $\ell_\pi^{(2)}$ (the diffractive scale of the host galaxy's ISM), cannot be located within a distance of approximately $10^{18}$\,cm from the source. If the close-in screen is situated farther away, the scintillation pattern it imposes on the FRB spectrum can be utilized to distinguish between the two classes of FRB models. This is due to the diffractive scale of such a screen, assuming it is an ionized SNR, is (eq. \ref{lpi})
\begin{equation}
  \ell_\pi^{(1)} \sim (2{\rm x10^7\, cm)}\, m^{-6/5} D_{1,18}^{17/5} \nu_9^{6/5} (\ell_{\rm max}^{(1)}/D_1)^{2/5}.
\end{equation}
which lies between the expected source sizes for the near-field and far-away FRB models, and the scintillation coherence bandwidth for this screen
\begin{equation}
    \delta\nu_{\rm sc}^{(1)} \sim \left[ {\ell_\pi^{(1)}\over \lambda}\right]^2 {2c\over \pi D_1} \sim {(4\,\rm kHz}) \, m^{-{12\over 5}} \nu_9^{22\over 5} D_{1,18}^{29\over 5} (\ell_{\rm max}^{(1)}/D_1)^{4\over 5}.
\end{equation}
is smaller than the MW-ISM value by several orders of magnitude, making it useful for distinguishing between different FRB models through the measurement of the associated modulation index.

\begin{figure}
    \centering
    \includegraphics[width=0.48\textwidth]{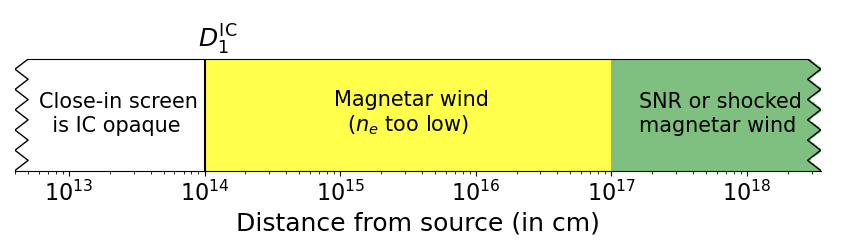}
    \caption{This figure shows the viability of a scattering screen at different distances from the FRB source that might magnify the source size and compromise the ability of the FRB host galaxy's ISM to differentiate between different FRB models. There are three different regions, coded by different colors, each with distinct properties. 
(a) \textit{White}: In this region, a screen capable of interfering with the resolving power of the ISM screen in the FRB host galaxy would not allow FRB radio waves to pass through due to the large induced Compton (IC) optical depth. Hence, such a screen cannot exist at distances smaller than approximately $\sim 10^{14}$ cm from the source.
(b) \textit{Yellow}: The density required for a screen in this region is larger by several orders of magnitude than what can be reasonably expected from the magnetar wind (see Eq. \ref{eq:Mdottwoscreen}).
(c) Green: A scattering screen in this region with the required density (Eq. \ref{ne-lim2}) is possible. This screen might be due to the presence of a supernova remnant or a shocked magnetar wind. However, in most cases, the scintillation from this close-in screen can be used to differentiate between different FRB models.
This figure uses $\xi=1$, meaning a point source has been broadened to the diffractive length scale of the host galaxy ISM screen due to strong scattering in the close-in screen.
    }
    \label{fig:dist-limits}
\end{figure}

To summarize, the analysis in this sub-section shows that the ability of a scattering screen within about one pc of the FRB source to magnify the source size and restrict the resolving power of the host galaxy's ISM scintillation is severely limited, as shown in Fig. \ref{fig:dist-limits}. This limitation primarily arises from two key requirements. First, the first screen must be optically thin to induced Compton scatterings while also broadening the effective size of the source to the extent that it suppresses strong frequency modulations by the second screen. The IC optical depth precludes a scattering screen within $\sim 10^{14}$cm that could restrict the resolving power of the second screen. Second, plausible sources of plasma for the first screen near the source are severely limited by the required electron density in the screen. For instance, the density for an electron-positron wind from a magnetar falls short by a few orders of magnitude at distances larger than $\sim 10^{14}$cm, where the screen becomes transparent to the IC. The considerations of free-free absorption and an upper limit to the electron density of the screen, which should be bounded by DM$_{\rm host}$, leave open the possibility of a screen at a distance of about 1 pc from the source. This screen is likely composed of a supernova remnant and shock-compressed magnetar wind, which could interfere with the host galaxy ISM screen.

\subsection{Scintillation in the Milky Way}
\label{sec:MW-scintillation}

The ISM of our galaxy causes scintillation of compact radio sources at low enough frequencies where the plasma density variations can induce sufficiently large electromagnetic phase variations.   For pulsars and FRBs,  scintillations are predominantly diffractive in nature combined with weaker refractive variations.  For most active galactic nuclei, only refractive scintillations are seen. The MW-ISM scintillation is not useful for constraining the FRB source size, as mentioned earlier, but they have been used to constrain the sizes of Galactic pulsar emission regions. In fact, it can cause significant flux variation over frequency scales greater than the scintillation bandwidth in the FRB host galaxy, which could affect the determination of the source size. The MW-ISM contribution to the flux variation over frequency scales of several times the scintillation bandwidth of the host galaxy needs to be subtracted to obtain the FRB source size. This subsection and the next one are devoted to quantifying how MW-ISM scintillation affects the amplitude of flux variation over frequency scales of several times the scintillation bandwidth in the host galaxy.

The scattering broadening time for the ISM plasma in our galaxy has been determined from pulsar surveys at different Galactic latitudes and longitudes \citep{CL2002},
and the mean scattering time for different latitudes ($b$) is approximately given by the following equation \citep{cordes2019}
\begin{equation}
   \delta t_{\rm s,G} \sim (4{\rm x 10^{-8}s}) \; |\sin b|^{-6/5}\nu_9^{-4.4}.
\end{equation}
From this we can quantify the scintillation coherence bandwidth for the Milky Way galaxy
\begin{equation}
    \delta\nu_{\rm sc,G} \approx {1\over 2\pi \delta t_{\rm s,G}} \sim ({\rm 4\, MHz})\; |\sin b|^{6/5}\nu_9^{4.4}.
\end{equation}

These scalings are not a substitute for the more accurate scattering information provided by the NE2001 model \citep{CL2002}, which should be used in analyzing FRB data. In fact, these scalings differ by more than order of magnitude at low galactic latitudes below $20^{\circ}$. The MW-ISM scintillation can affect the amplitude of flux variation across frequency scales greater than the scintillation bandwidth for the host galaxy, and that is estimated in the sub-subsection below.

\subsubsection{Frequency dependence of flux for a scintillating source}
\label{sec:nudepscint}
Consider a point source which is undergoing strong scintillation at some $\nu_0$.
At such a frequency, the phase in eq. \ref{eq:Ai} is highly oscillatory and contribution to the integral comes from the regions in which the phase is extremal. The separation between such regions in the scattering screen is given by $\ell_{\pi}$ , and the visible size of the screen is $R_{\rm sc}\approx R_F^2/\ell_{\pi}$, implying there are $N_{\rm sc}(\nu_0)\approx (R_{\rm sc}/\ell_{\pi})^2\approx (\nu_0/\delta \nu_{\rm sc})^2$ regions that contribute to the flux. As a result, the flux can be compared to a multi-slit experiment in which we have $N_{\rm sc}(\nu_0)$ slits, each contributing a comparable amplitude with a random phase, i.e. the wave amplitude is proportional to
\begin{equation}
	A(\nu_0)\propto \sum_i^{N_{\rm sc}(\nu_0)} \cos[\phi_{p,i}^{\nu_0}+\phi_{g,i}^{\nu_0}].
\end{equation}
where $\phi_{g,i}=x_{i}^2/R_{F}^2$ is the geometrical phase shift for the i-th slit at a 
distance $x_{i}$ from the screen-center as measured relative to the line going directly through the screen.
The root mean square plasma phase shift for the same slit is given by the phase structure function, 
$\sqrt{\langle \phi_{p,i}^2\rangle}\sim (x_{i}/\ell_{\pi})^{5/6}$. The distance between two random slits is of order $R_{\rm sc}$, and therefore the geometrical phase difference  is $\sim R_{\rm sc}/\ell_{\pi}\sim \nu_0/\delta \nu_{\rm sc}$, i.e. the geometrical phase difference (slightly) dominates over the plasma phase difference. We therefore consider only the geometrical phase
shift for the estimates we discuss below.

Consider a frequency range $\delta\nu_{\rm ob} \ll \nu_0$ in the vicinity of $\nu_0$. The small shift in frequency relative to $\nu_0$ leads to a small shift of the slit positions on the screen as well as to a small change in the number of slits (some slits appear/disappear compared to the case of frequency $\nu_0$). 
We consider first the phase shift in each slit:  $\Delta \phi_{g,i}\equiv \lvert \phi_{g,i}^{\nu}-\phi_{g,i}^{\nu_0}\rvert \sim \pi \delta\nu_{\rm ob}/\delta \nu_{\rm sc}$.  Consequently, we see that for $\delta\nu_{\rm ob}=\delta \nu_{\rm sc}$, $\Delta  \phi_{g,i}\sim \pi$, i.e. all slits have changed their phases randomly and that as a result we have $\delta f/f \sim 1$ over the frequency bandwidth of $\delta\nu_{\rm sc}$ at $\nu_0$. 

Similarly, $N_{\rm sc}$ has a power-law dependence on $\nu_0$. 
Therefore, the change in the number of slits at $\nu$ compared to $\nu_0$, $\Delta N_{\rm sc}=|N_{\rm sc}(\nu)-N_{\rm sc}(\nu_0)|\approx N_{\rm sc}(\nu_0) \delta\nu_{\rm ob}/\nu_0 \ll N_{\rm sc}(\nu_0)$.

We can now calculate how the wave amplitude changes with $\nu$ -- $A(\nu)\propto \sum_i^{N_{\rm sc}(\nu)} \cos[\phi_{g,i}^{\nu}]$ -- in the neighborhood of $\nu_0$,
\begin{equation}
	\frac{|A(\nu)-A(\nu_0)|}{|A(\nu_0)|} \approx \frac{|\sum_i^{N_{\rm sc}(\nu)}\sin[\phi_{g,i}^{\nu_0}]\Delta \phi_{g,i}|}{|\sum_i^{N_{\rm sc}(\nu_0)}\cos[\phi_{g,i}^{\nu_0}]|}\approx C_A \frac{\delta\nu_{\rm ob}}{\delta \nu_{\rm sc}}.
\end{equation}
where $C_A$ is a constant of order unity.
The flux difference is given by
\begin{equation}
	\frac{|f(\nu)-f(\nu_0)|}{|f(\nu_0)|} \approx 2\frac{|\lvert A(\nu)\rvert-\lvert A(\nu_0)\rvert|}{|A(\nu_0)|}
	\approx 2C_A \frac{\delta\nu_{\rm ob}}{\delta \nu_{\rm sc}}.
\end{equation}
The conclusion is that for $\delta\nu_{\rm ob} \ll \delta \nu_{\rm sc}$, the flux at $\nu_0-\delta\nu_{\rm ob} < \nu<\nu_0+\delta\nu_{\rm ob}$ varies by a small amount, $\delta f/f\sim \delta\nu_{\rm ob}/\delta \nu_{\rm sc}$.

The implication of this calculation is that if we consider a source that is strongly modulated by two plasma scattering screens, one in the host galaxy, and the other in the MW-ISM, such that, for example, $\delta \nu_{\rm sc,h}/\delta \nu_{\rm sc,MW}\ll 1$ (note that this ratio is approximately independent of the central observed frequency as $\delta \nu_{\rm sc}$ from both screens carries the same wavelength dependence), then on a frequency scale $\delta\nu_{\rm ob} \sim \nu_{\rm sc,h}$, 
the flux modulation due to the host screen is $\delta f/f\sim 1$ while due to the MW-ISM screen 
the flux variation on over the same frequency range ($\nu_{\rm sc,h}$) is much smaller than unity.
In other words, on such small frequency separations, significant flux fluctuations are completely dominant by the host plasma screen with the lower scintillation bandwidth. Therefore, the presence or absence of such fluctuations for an FRB which is known to be scatter broadened by a plasma screen within its host galaxy can be used to place a limit on the FRB source size.

As a numerical example consider the situation in which at 1GHz,  $\delta \nu_{\rm sc,h}\sim1$\,kHz and $\delta \nu_{\rm sc,MW}\sim1$\,MHz; The presence of the host screen can be securely confirmed if we observe scatter broadening at a level of $\delta t_s\sim 1$\,ms. 
At an observed frequency of 3GHz, we have $\delta \nu_{\rm sc,h}\sim0.1$\,MHz and 
$\delta \nu_{\rm sc,MW}\sim0.1$\,GHz. If we detect strong flux modulation on a scale of 
$0.1$\,MHz, then that can only be due a scattering screen in the FRB host galaxy when 
the source size is of order of or less than $\ell_\pi$, see \S \ref{sec:main-idea}) as the modulation from the MW-ISM screen is suppressed to a level of $\delta f/f\sim 10^{-3}\ll 1$.

\subsection{Level of flux fluctuations - general case} \label{sec:fluxfluctuations}
As described in the previous subsections, the extent to which the flux is modulated with frequency by a scintillating screen within the FRB's host galaxy, depends\footnote{ For clarity, we assume here the situation where the honeycomb instability does not significantly affect the FRB waves (see \S \ref{fragmentation-instability} for details).} on (i) the transverse source size, (ii) the scintillation regime (strong or weak) and (iii) the spectral resolution of the detector. 
(i) is described in \S \ref{sec:intro}, and can be approximated as $\delta f/f\sim \min(1, \ell_{\pi}/R_{\rm s})$ or equivalently $\delta f/f\sim \min[1, (\nu/\nu_{\rm so})^{6/5}]$. (ii) is described in \cite{Narayan1992} and is given by $\delta f/f\sim \min[1, (R_{\rm F}/\ell_{\pi})^{5/6}]$ or equivalently $\delta f/f\sim \min[1, (\nu/\delta \nu_{\rm sc})^{5/12}]$. Finally, (iii) is described in \S \ref{section:source-size} for the case $\delta \nu_{\rm ob}\gg \delta \nu_{\rm sc}$ and in \S \ref{sec:nudepscint} for the case $\delta \nu_{\rm sc}\gg \delta \nu_{\rm ob}$ and for the general case is then $\delta f/f\sim \min[(\delta \nu_{\rm sc}/\delta \nu_{\rm ob})^{1/2},(\delta \nu_{\rm ob}/\delta \nu_{\rm sc})]$. These effects add together linearly, leading to a general expression for $\delta f/f$,
\begin{equation}
    \frac{\delta f}{f}\!\approx\! \min\left[1, \left(\frac{\nu}{\nu_{\rm so}}\right)^{6/5}\right]\! \min\left[1, \left(\frac{\nu}{\delta \nu_{\rm sc}}\right)^{5/12}\right] \!\min\left[\frac{\delta \nu_{\rm ob}}{\delta \nu_{\rm sc}},\left(\frac{\delta \nu_{\rm sc}}{\delta \nu_{\rm ob}}\right)^{1/2}\right]
\end{equation}
This is demonstrated in figure \ref{fig:delfoverf}. In particular, we see that significant flux modulations are only possible if the emission site radius is small as described in \S \ref{sec:intro}. The large, $\gtrsim 3\times 10^4$ separation of scales between the typical distances of the emission site from the central engine in magnetospheric and blast-wave models, makes the test of flux modulation on the frequency scale of $(2\pi\delta t_s)^{-1}$ a strong candidate for distinguishing between the model classes. However, we caution that the value of $\ell_{\pi}$ (which is the x-axis of figure \ref{fig:delfoverf}) cannot be uniquely determined from observations (see eq. \ref{lpi2} - there is uncertainty involving the exact value of $d_L$) and as a result by measuring $\delta f/f$, it will generally not be possible to rule out both values above the minimum blast-wave distance and below the maximum magnetospheric distance. Nonetheless, this ambiguity goes only in one direction: if $\delta f/f$ is small then the source distance cannot be strongly constrained without additional independent determination of $\ell_{\pi}$, while if $\delta f/f$ is of order unity, this necessitates a small source distance, and in addition constrains $\ell_{\pi}$ and teaches us about the host-galaxy scattering screen.

\begin{figure}
\includegraphics[width = 0.53\textwidth]{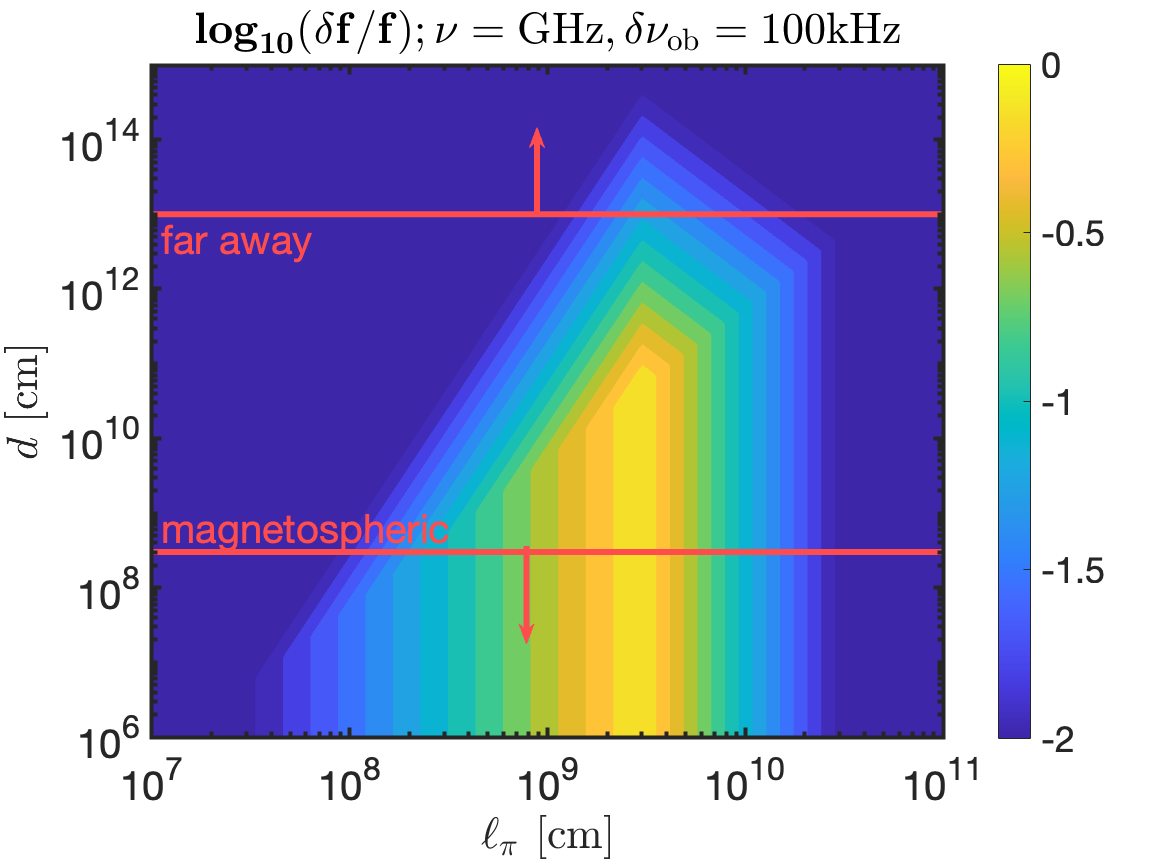}
\caption{Level of flux modulation, $\delta f/f$, for an FRB source whose radiation is passing through a scintillation screen in the host galaxy. Results are plotted as a function of the diffractive scale in the scintillating screen, $\ell_{\pi}$ and the distance of the FRB source from the NS, $d$. In far-away models, we focus on blast waves for which Eq. \ref{eq:Rs} can be used to relate $d$ to the lateral source size, $R_{s}$, while in magnetospheric models the same relation provides an upper limit on $R_{s}$ (corresponding to a lower limit on $\delta f/f$). We also consider an observed frequency of $\nu=1$\,GHz, a detector with spectral resolution $\delta \nu_{\rm ob}=100$\,kHz, a scintillation screen at a distance of $1$\,kpc from the FRB source and an FRB duration of $t_{\rm FRB}=3$\,ms. Overall, significant flux modulations are only possible if the emission site radius is small. The typical distance of the FRB source assuming maser emission in blast wave models is $d\gtrsim 10^{13}$cm, which is constrained by the requirement that the medium must be transparent to induced-Compton scatterings (Eq. \ref{eq-IC}) while still being to reproduce the characteristic timescale, frequency and energy of an FRB \citep{Margalit+20,LKZ2020}.}
    \label{fig:delfoverf}
\end{figure} 

\section{FRB data: scintillation bandwidth amplitude}

FRBs show a wide variety of radio spectra that include contributions from the source, from propagation through plasma in the host galaxy, including through the circumsource region and the MW-ISM.   As discussed in
\citet{Cordes2022},   neither the intergalactic medium nor circumgalactic media appear to cause multipath propagation that would lead to constructive and destructive interference vs. frequency. 

The salient features of FRB spectra include (1) spectral islands that  occur at lower frequencies vs. time through the burst (the `sad trombone' effect; \citealt{Hessels+19}); (2)  in some cases, an increasing asymmetry  of the burst at lower frequencies indicative of multipath scattering broadening; in the vast majority  of cases this scattering occurs in the host galaxy;
and (3) frequency structure from interstellar scattering from the Galaxy \citep[e.g.][and references therein]{2021ApJ...911..102O}.     A suggested trend is that repeating FRBs tend to show spectral islands while 
non-(or not yet) repeating FRBs do not \citep{Pleunis2021b}.   

Some FRBs show neither scatter-broadening nor Galactic (MW-ISM) scintillation.   
The absence of scatter broadening implies that scattering in the host galaxy is weak enough that it would be manifested instead as  frequency structure, with scintillation bandwidths that  could range from tens to hundreds
of MHz down to kHz scales.   For these cases,  the presence or absence of scintillation structure could yield a constraint on the source size, as discussed here.   However, it might also be masked by spectral islands associated with the emission process or confused by plasma lensing. A case in point is FRB~20121102A which shows MW-ISM scintillation but without a hint of any scattering broadening from the dwarf host galaxy
\citep[][]{Gajjar2018, 2021ApJ...911..102O}.    This FRB shows prominent spectral islands but no additional frequency structure that could unequivocally be associated with scattering in the host galaxy. 

MW-ISM scintillation has been seen  in only  a few FRBs.  Selection effects undoubtedly play a prominent role because adequate signal to noise ratio is needed to measure and quantify the scintillations.   Future work with more sensitive instruments can alleviate this issue and perhaps identify FRBs where MW-ISM scintillations are absent because of scattering in the host galaxy.

\section{Summary}

Almost all the models that have been suggested for FRB coherent radiation fall into two broad classes: near-field models (where radiation is produced within the magnetosphere of a magnetar) and far-away models that posit coherent radio waves are produced by the dissipation of an ultra-relativistic jet outside the magnetosphere. The expected source sizes for these two classes of models are $\lta10^7$ cm and $\gta10^{9}$  cm, respectively. It turns out that the diffractive scale for strong scintillation at 1 GHz, for scattering in the host galaxy ISM when the scatter broadening is 0.1 ms, is of the order of a few times 10$^8$ cm (eq. \ref{lpi2}). This scale is conveniently between the source sizes for the two classes of models and well suited for determining which of these models is responsible for the FRB radiation. This can be achieved by determining the modulation index for the host galaxy scintillation using high-frequency resolution data of order a few kHz.

As primary targets for this search we suggest a two-step analysis. The first step is to identify FRBs that have host-galaxy-dominated scattering widths. Next, for these bursts, an ACF analysis should be performed to quantify the modulation index corresponding to the scintillation width of $(2\pi\delta t_s)^{-1}$; where $\delta t_s$ is the scatter broadening time-scale. FRBs with no observed scatter broadening can also be useful for this analysis, with the following caveats. The lack of observed scatter broadening might be because either (i) their scatter broadening time is shorter than their intrinsic variability time-scale or the instrument's temporal resolution or (ii) because scatter broadening by plasma in their host galaxy is too small ($\mu$s or less) to distinguish it from the MW-ISM scintillation pattern. While the second case is of little interest for the purpose of this work, in the first case, the corresponding scintillation bandwidth would be relatively large. This will allow it to be more readily resolved (than for FRBs with observed, longer, scatter broadening times) and make testing for frequency modulations on this scale technically easier. The disadvantage of working with such bursts is that one does not know apriori on which frequency scale to search for modulation, thereby making it hard to interpret when no flux fluctuations are found in the data. Nonetheless, if order unity flux modulations are seen on some frequency scale, this immediately limits the lateral size of the source to be less than $\ell_{\pi}$ (which in turn is directly related to the scintillation bandwidth, see eq. \ref{lpi2}) for the plasma screen in the host galaxy.

The scattering bandwidth is a rapidly increasing function of frequency (approximately $\delta \nu_{sc}\propto \nu^{4.4}$, see e.g. \citealt{BK2020}). Thus, if an FRB is observed over a wide frequency band, scatter broadening might be easier to identify at lower frequencies, where it is more pronounced, but frequency modulations could be easier to search for towards the higher end of the observed frequency band. This approach significantly relaxes the required frequency resolution needed to search for flux modulations, but should be used with caution, considering that at sufficiently high frequencies, the radiation will transition to the weak scattering regime, causing a reduction in the flux modulations, even for a source that is effectively point-like (see Eq. \ref{eq:nustar} and \S \ref{sec:fluxfluctuations}). 

We note that most FRBs have currently been detected by CHIME in the frequency band of 400-800 MHz. While the analysis suggested in this work might be more challenging to carry out in this range, the limited data at 1.4 GHz suggests that the FRB rate at this higher frequency is similar to the CHIME rate, making the scintillation analysis easier.

If the frequency resolution of an FRB detector ($\delta\nu_{\rm ob}$) is larger than the scintillation bandwidth for the host galaxy ($\delta\nu_{\rm sc}$), then the amplitude of flux variations due to scintillation is suppressed, even for a point source, by a factor of $\sim (\delta\nu_{\rm ob}/\delta\nu_{\rm sc})^{1/2}$ \citep{BKN2022}. This suppression occurs because the bin size of the detector's frequency channel ($\delta\nu_{\rm ob}$) is larger than the scintillation bandwidth. Therefore, fluctuations of order unity across $\delta \nu_{\rm sc}$ are reduced due to the averaging. Consequently, the upper limit that can be placed on the source size is weakened to $R_{s}<\ell_\pi (\delta\nu_{\rm ob}/\delta\nu_{\rm sc})^{1/2}$.

We discussed various propagation effects in \S\ref{fragmentation-instability} \& \S\ref{sec:effects-on-ACF} that could influence the modulation amplitude and potentially complicate the ability to place a limit on the source size. One of the more serious concerns is the fragmentation of the FRB pulse within a fraction of a pc of the source, which can mimic certain aspects of scintillation. However, we found that this possibility, along with the others we have looked into, does not compromise the ability of the ACF analysis and modulation index to estimate the source size.

There are several examples of FRBs with scattering widths known to be caused by turbulent plasma in the host galaxy. \cite{Cordes2022} considered a sample of 14 FRBs with redshifts. Scattering widths were measured for 9 of these FRBs, many of which exhibited scatter broadening in excess of what one expects for the Milky Way ISM. Of particular note are FRBs 20181112A, 20190102B, and 20190611B, which have scattering widths of 0.02, 0.04, and 0.18 ms, respectively. The corresponding scintillation bandwidths are 8, 4, and 0.9 kHz. The first two of these, i.e., FRBs 20181112A and 20190102B, have scintillation bandwidths that could have been investigated by some of the existing FRB surveys such as CHIME, and an auto-correlation analysis carried out to determine the scintillation index ($m_I$). Similarly, \cite{Sammons2023} recently analyzed 10 FRBs, and for three of these (FRBs 20190608B, 20210320C, \& 20201124A) they make a compelling case that their scattering widths are due to plasma screens in their host galaxies. FRBs like these which have scatter broadening dominated by the host galaxy are good candidates for ACF analysis that can decide between the different classes of proposed FRB mechanisms. 

\bigskip\bigskip
\noindent {\bf ACKNOWLEDGEMENTS}

\medskip
We are grateful for excellent discussions with Robert Main, Kiyoshi Masui, Daniele Michilli, Kenzie Nimmo, Ziggy Pleunis, Jason Hessels and his group in Amsterdam. We are indebted to Jason Hessels for reading the draft of the paper and providing numerous suggestions to improve the presentation, as well as for pointing out several papers to cite that we had missed. PK thanks Ue-Li Pen for useful discussions about using scintillation to study the radio pulsar mechanism. We are highly indebted to the referee for numerous suggestions to improve the readability of the paper.
PB was supported by a grant (no. 2020747) from the United States-Israel Binational Science Foundation (BSF), Jerusalem, Israel. PK's work was funded in part by an NSF grant AST-2009619. OG was supported through a fellowship from the Graduate School, UT Austin.
JMC is supported by the NANOGrav Physics Frontiers Center, which  receives support from National Science Foundation (NSF) Physics Frontiers Center award numbers 1430284 and 2020265,
and by the National Aeronautics and Space Administration (NASA 80NSSC20K0784). 
We used ChatGPT for checking the grammar and for simplifying some long sentences to express our ideas more concisely. However, the paper was not written by ChatGPT.

\medskip

\noindent {\bf DATA AVAILABILITY}

\medskip
The code developed to perform calculations in this paper is available upon request.
\bibliography{main_submitted}

\begin{thebibliography}{}
\makeatletter
\relax
\def\mn@urlcharsother{\let\do\@makeother \do\$\do\&\do\#\do\^\do\_\do\%\do\~}
\def\mn@doi{\begingroup\mn@urlcharsother \@ifnextchar [ {\mn@doi@}
  {\mn@doi@[]}}
\def\mn@doi@[#1]#2{\def\@tempa{#1}\ifx\@tempa\@empty \href
  {http://dx.doi.org/#2} {doi:#2}\else \href {http://dx.doi.org/#2} {#1}\fi
  \endgroup}
\def\mn@eprint#1#2{\mn@eprint@#1:#2::\@nil}
\def\mn@eprint@arXiv#1{\href {http://arxiv.org/abs/#1} {{\tt arXiv:#1}}}
\def\mn@eprint@dblp#1{\href {http://dblp.uni-trier.de/rec/bibtex/#1.xml}
  {dblp:#1}}
\def\mn@eprint@#1:#2:#3:#4\@nil{\def\@tempa {#1}\def\@tempb {#2}\def\@tempc
  {#3}\ifx \@tempc \@empty \let \@tempc \@tempb \let \@tempb \@tempa \fi \ifx
  \@tempb \@empty \def\@tempb {arXiv}\fi \@ifundefined
  {mn@eprint@\@tempb}{\@tempb:\@tempc}{\expandafter \expandafter \csname
  mn@eprint@\@tempb\endcsname \expandafter{\@tempc}}}

\bibitem[\protect\citeauthoryear{{Backer}}{{Backer}}{1974}]{1974ApJ...190..667B}
{Backer} D.~C.,  1974, \mn@doi [\apj] {10.1086/152924}, \href
  {https://ui.adsabs.harvard.edu/abs/1974ApJ...190..667B} {190, 667}

\bibitem[\protect\citeauthoryear{{Beloborodov}}{{Beloborodov}}{2017}]{Beloborodov17}
{Beloborodov} A.~M.,  2017, \mn@doi [\apjl] {10.3847/2041-8213/aa78f3}, \href
  {https://ui.adsabs.harvard.edu/abs/2017ApJ...843L..26B} {843, L26}

\bibitem[\protect\citeauthoryear{{Beloborodov}}{{Beloborodov}}{2019}]{Beloborodov19}
{Beloborodov} A.~M.,  2019, arXiv e-prints, \href
  {https://ui.adsabs.harvard.edu/abs/2019arXiv190807743B} {p. arXiv:1908.07743}

\bibitem[\protect\citeauthoryear{{Beloborodov}}{{Beloborodov}}{2021}]{Beloborodov2021}
{Beloborodov} A.~M.,  2021, \mn@doi [\apjl] {10.3847/2041-8213/ac2fa0}, \href
  {https://ui.adsabs.harvard.edu/abs/2021ApJ...922L...7B} {922, L7}

\bibitem[\protect\citeauthoryear{{Beniamini} \& {Kumar}}{{Beniamini} \&
  {Kumar}}{2020}]{BK2020}
{Beniamini} P.,  {Kumar} P.,  2020, \mn@doi [\mnras] {10.1093/mnras/staa2489},
  \href {https://ui.adsabs.harvard.edu/abs/2020MNRAS.498..651B} {498, 651}

\bibitem[\protect\citeauthoryear{{Beniamini} \& {Kumar}}{{Beniamini} \&
  {Kumar}}{2023}]{BK2023}
{Beniamini} P.,  {Kumar} P.,  2023, \mn@doi [\mnras] {10.1093/mnras/stad028},
  \href {https://ui.adsabs.harvard.edu/abs/2023MNRAS.519.5345B} {519, 5345}

\bibitem[\protect\citeauthoryear{{Beniamini}, {Kumar}  \&
  {Narayan}}{{Beniamini} et~al.}{2022}]{BKN2022}
{Beniamini} P.,  {Kumar} P.,   {Narayan} R.,  2022, \mn@doi [\mnras]
  {10.1093/mnras/stab3730}, \href
  {https://ui.adsabs.harvard.edu/abs/2022MNRAS.510.4654B} {510, 4654}

\bibitem[\protect\citeauthoryear{{Bhardwaj} et~al.,}{{Bhardwaj}
  et~al.}{2021}]{Bhardwaj2021}
{Bhardwaj} M.,  et~al., 2021, \mn@doi [\apjl] {10.3847/2041-8213/abeaa6}, \href
  {https://ui.adsabs.harvard.edu/abs/2021ApJ...910L..18B} {910, L18}

\bibitem[\protect\citeauthoryear{{Bochenek}, {Ravi}, {Belov}, {Hallinan},
  {Kocz}, {Kulkarni}  \& {McKenna}}{{Bochenek} et~al.}{2020}]{STARE2020}
{Bochenek} C.~D.,  {Ravi} V.,  {Belov} K.~V.,  {Hallinan} G.,  {Kocz} J.,
  {Kulkarni} S.~R.,   {McKenna} D.~L.,  2020, \mn@doi [\nat]
  {10.1038/s41586-020-2872-x}, \href
  {https://ui.adsabs.harvard.edu/abs/2020Natur.587...59B} {587, 59}

\bibitem[\protect\citeauthoryear{{CHIME/FRB Collaboration} et~al.,}{{CHIME/FRB
  Collaboration} et~al.}{2018}]{CHIME2018}
{CHIME/FRB Collaboration} et~al., 2018, \mn@doi [\apj]
  {10.3847/1538-4357/aad188}, \href
  {https://ui.adsabs.harvard.edu/abs/2018ApJ...863...48C} {863, 48}

\bibitem[\protect\citeauthoryear{{Chatterjee} et~al.,}{{Chatterjee}
  et~al.}{2017}]{Chatterjee+17}
{Chatterjee} S.,  et~al., 2017, \mn@doi [\nat] {10.1038/nature20797}, \href
  {https://ui.adsabs.harvard.edu/abs/2017Natur.541...58C} {541, 58}

\bibitem[\protect\citeauthoryear{{Chime/Frb Collaboration} et~al.,}{{Chime/Frb
  Collaboration} et~al.}{2020}]{CHIME+20}
{Chime/Frb Collaboration} et~al., 2020, \mn@doi [\nat]
  {10.1038/s41586-020-2398-2}, \href
  {https://ui.adsabs.harvard.edu/abs/2020Natur.582..351C} {582, 351}

\bibitem[\protect\citeauthoryear{{Chime/Frb Collaboration} Andersen
  et~al.,}{{Chime/Frb Collaboration} et~al.}{2022}]{subsecPCHIME}
{Chime/Frb Collaboration} Andersen B.~C.,  et~al., 2022, \mn@doi [\nat]
  {10.1038/s41586-022-04841-8}, \href
  {https://ui.adsabs.harvard.edu/abs/2022Natur.607..256C} {607, 256}

\bibitem[\protect\citeauthoryear{{Cordes}}{{Cordes}}{2000}]{jmc2000}
{Cordes} J.~M.,  2000, \mn@doi [arXiv e-prints]
  {10.48550/arXiv.astro-ph/0007231}, \href
  {https://ui.adsabs.harvard.edu/abs/2000astro.ph..7231C} {pp
  astro--ph/0007231}

\bibitem[\protect\citeauthoryear{{Cordes} \& {Chatterjee}}{{Cordes} \&
  {Chatterjee}}{2019}]{cordes2019}
{Cordes} J.~M.,  {Chatterjee} S.,  2019, \mn@doi [\araa]
  {10.1146/annurev-astro-091918-104501}, \href
  {https://ui.adsabs.harvard.edu/abs/2019ARA&A..57..417C} {57, 417}

\bibitem[\protect\citeauthoryear{{Cordes} \& {Lazio}}{{Cordes} \&
  {Lazio}}{2002}]{CL2002}
{Cordes} J.~M.,  {Lazio} T.~J.~W.,  2002, arXiv e-prints, \href
  {https://ui.adsabs.harvard.edu/abs/2002astro.ph..7156C} {pp
  astro--ph/0207156}

\bibitem[\protect\citeauthoryear{{Cordes} \& {Rickett}}{{Cordes} \&
  {Rickett}}{1998}]{Cordes1998}
{Cordes} J.~M.,  {Rickett} B.~J.,  1998, \mn@doi [\apj] {10.1086/306358}, \href
  {https://ui.adsabs.harvard.edu/abs/1998ApJ...507..846C} {507, 846}

\bibitem[\protect\citeauthoryear{{Cordes} \& {Wasserman}}{{Cordes} \&
  {Wasserman}}{2016}]{CW2016}
{Cordes} J.~M.,  {Wasserman} I.,  2016, \mn@doi [\mnras]
  {10.1093/mnras/stv2948}, \href
  {https://ui.adsabs.harvard.edu/abs/2016MNRAS.457..232C} {457, 232}

\bibitem[\protect\citeauthoryear{Cordes, Weisberg  \& Boriakoff}{Cordes
  et~al.}{1983}]{cwb83}
Cordes J.~M.,  Weisberg J.~M.,   Boriakoff V.,  1983, apj, 268, 370

\bibitem[\protect\citeauthoryear{{Cordes}, {Bhat}, {Hankins}, {McLaughlin}  \&
  {Kern}}{{Cordes} et~al.}{2004}]{2004ApJ...612..375C}
{Cordes} J.~M.,  {Bhat} N.~D.~R.,  {Hankins} T.~H.,  {McLaughlin} M.~A.,
  {Kern} J.,  2004, \mn@doi [\apj] {10.1086/422495}, \href
  {https://ui.adsabs.harvard.edu/abs/2004ApJ...612..375C} {612, 375}

\bibitem[\protect\citeauthoryear{{Cordes}, {Ocker}  \& {Chatterjee}}{{Cordes}
  et~al.}{2022}]{Cordes2022}
{Cordes} J.~M.,  {Ocker} S.~K.,   {Chatterjee} S.,  2022, \mn@doi [\apj]
  {10.3847/1538-4357/ac6873}, \href
  {https://ui.adsabs.harvard.edu/abs/2022ApJ...931...88C} {931, 88}

\bibitem[\protect\citeauthoryear{{Driessen}, {Janssen}, {Bassa}, {Stappers}  \&
  {Stinebring}}{{Driessen} et~al.}{2019}]{Driessen2019}
{Driessen} L.~N.,  {Janssen} G.~H.,  {Bassa} C.~G.,  {Stappers} B.~W.,
  {Stinebring} D.~R.,  2019, \mn@doi [\mnras] {10.1093/mnras/sty3192}, \href
  {https://ui.adsabs.harvard.edu/abs/2019MNRAS.483.1224D} {483, 1224}

\bibitem[\protect\citeauthoryear{{Farah} et~al.,}{{Farah}
  et~al.}{2018}]{Farah2018}
{Farah} W.,  et~al., 2018, \mn@doi [\mnras] {10.1093/mnras/sty1122}, \href
  {https://ui.adsabs.harvard.edu/abs/2018MNRAS.478.1209F} {478, 1209}

\bibitem[\protect\citeauthoryear{{Gajjar} et~al.,}{{Gajjar}
  et~al.}{2018}]{Gajjar2018}
{Gajjar} V.,  et~al., 2018, \mn@doi [\apj] {10.3847/1538-4357/aad005}, \href
  {https://ui.adsabs.harvard.edu/abs/2018ApJ...863....2G} {863, 2}

\bibitem[\protect\citeauthoryear{{Gopinath} et~al.,}{{Gopinath}
  et~al.}{2023}]{Gopinath2023}
{Gopinath} A.,  et~al., 2023, \mn@doi [arXiv e-prints]
  {10.48550/arXiv.2305.06393}, \href
  {https://ui.adsabs.harvard.edu/abs/2023arXiv230506393G} {p. arXiv:2305.06393}

\bibitem[\protect\citeauthoryear{{Gwinn} et~al.,}{{Gwinn}
  et~al.}{1997}]{Gwinn1997}
{Gwinn} C.~R.,  et~al., 1997, \mn@doi [\apjl] {10.1086/310734}, \href
  {https://ui.adsabs.harvard.edu/abs/1997ApJ...483L..53G} {483, L53}

\bibitem[\protect\citeauthoryear{{Gwinn} et~al.,}{{Gwinn}
  et~al.}{2012}]{Gwinn2012}
{Gwinn} C.~R.,  et~al., 2012, \mn@doi [\apj] {10.1088/0004-637X/758/1/7}, \href
  {https://ui.adsabs.harvard.edu/abs/2012ApJ...758....7G} {758, 7}

\bibitem[\protect\citeauthoryear{{Hessels} et~al.,}{{Hessels}
  et~al.}{2019}]{Hessels+19}
{Hessels} J.~W.~T.,  et~al., 2019, \mn@doi [\apjl] {10.3847/2041-8213/ab13ae},
  \href {https://ui.adsabs.harvard.edu/abs/2019ApJ...876L..23H} {876, L23}

\bibitem[\protect\citeauthoryear{{Hewitt} et~al.,}{{Hewitt}
  et~al.}{2023}]{Hewitt2023}
{Hewitt} D.~M.,  et~al., 2023, \mn@doi [arXiv e-prints]
  {10.48550/arXiv.2308.12118}, \href
  {https://ui.adsabs.harvard.edu/abs/2023arXiv230812118H} {p. arXiv:2308.12118}

\bibitem[\protect\citeauthoryear{{Johnson}, {Gwinn}  \& {Demorest}}{{Johnson}
  et~al.}{2012}]{Johnson2012}
{Johnson} M.~D.,  {Gwinn} C.~R.,   {Demorest} P.,  2012, \mn@doi [\apj]
  {10.1088/0004-637X/758/1/8}, \href
  {https://ui.adsabs.harvard.edu/abs/2012ApJ...758....8J} {758, 8}

\bibitem[\protect\citeauthoryear{{Kirsten} et~al.,}{{Kirsten}
  et~al.}{2022}]{Kirsten2022}
{Kirsten} F.,  et~al., 2022, \mn@doi [\nat] {10.1038/s41586-021-04354-w}, \href
  {https://ui.adsabs.harvard.edu/abs/2022Natur.602..585K} {602, 585}

\bibitem[\protect\citeauthoryear{{Kumar} \& {Bo{\v{s}}njak}}{{Kumar} \&
  {Bo{\v{s}}njak}}{2020}]{KumarBosnjak2020}
{Kumar} P.,  {Bo{\v{s}}njak} {\v{Z}}.,  2020, \mn@doi [\mnras]
  {10.1093/mnras/staa774}, \href
  {https://ui.adsabs.harvard.edu/abs/2020MNRAS.494.2385K} {494, 2385}

\bibitem[\protect\citeauthoryear{{Kumar} \& {Lu}}{{Kumar} \&
  {Lu}}{2020}]{KumarLu-IC-scat-2020}
{Kumar} P.,  {Lu} W.,  2020, \mn@doi [\mnras] {10.1093/mnras/staa801}, \href
  {https://ui.adsabs.harvard.edu/abs/2020MNRAS.494.1217K} {494, 1217}

\bibitem[\protect\citeauthoryear{{Kumar}, {Lu}  \& {Bhattacharya}}{{Kumar}
  et~al.}{2017}]{Kumar+17}
{Kumar} P.,  {Lu} W.,   {Bhattacharya} M.,  2017, \mn@doi [\mnras]
  {10.1093/mnras/stx665}, \href
  {https://ui.adsabs.harvard.edu/abs/2017MNRAS.468.2726K} {468, 2726}

\bibitem[\protect\citeauthoryear{{Lin}, {van Kerkwijk}, {Main}, {Mahajan},
  {Pen}  \& {Kirsten}}{{Lin} et~al.}{2023}]{Lin2023}
{Lin} R.,  {van Kerkwijk} M.~H.,  {Main} R.,  {Mahajan} N.,  {Pen} U.-L.,
  {Kirsten} F.,  2023, \mn@doi [\apj] {10.3847/1538-4357/acba95}, \href
  {https://ui.adsabs.harvard.edu/abs/2023ApJ...945..115L} {945, 115}

\bibitem[\protect\citeauthoryear{{Lorimer} \& {Kramer}}{{Lorimer} \&
  {Kramer}}{2004}]{Lorimer2004}
{Lorimer} D.~R.,  {Kramer} M.,  2004, {Handbook of Pulsar Astronomy}.
 Vol. 4

\bibitem[\protect\citeauthoryear{{Lorimer}, {Bailes}, {McLaughlin}, {Narkevic}
  \& {Crawford}}{{Lorimer} et~al.}{2007}]{Lorimer+07}
{Lorimer} D.~R.,  {Bailes} M.,  {McLaughlin} M.~A.,  {Narkevic} D.~J.,
  {Crawford} F.,  2007, \mn@doi [Science] {10.1126/science.1147532}, \href
  {https://ui.adsabs.harvard.edu/abs/2007Sci...318..777L} {318, 777}

\bibitem[\protect\citeauthoryear{{Lovelace}}{{Lovelace}}{1970}]{1970PhDT.......113L}
{Lovelace} R.~V.~E.,  1970, PhD thesis, -

\bibitem[\protect\citeauthoryear{{Lu}, {Kumar}  \& {Zhang}}{{Lu}
  et~al.}{2020}]{LKZ2020}
{Lu} W.,  {Kumar} P.,   {Zhang} B.,  2020, \mn@doi [\mnras]
  {10.1093/mnras/staa2450}, \href
  {https://ui.adsabs.harvard.edu/abs/2020MNRAS.498.1397L} {498, 1397}

\bibitem[\protect\citeauthoryear{{Lu}, {Beniamini}  \& {Kumar}}{{Lu}
  et~al.}{2022}]{LBK2022}
{Lu} W.,  {Beniamini} P.,   {Kumar} P.,  2022, \mn@doi [\mnras]
  {10.1093/mnras/stab3500}, \href
  {https://ui.adsabs.harvard.edu/abs/2022MNRAS.510.1867L} {510, 1867}

\bibitem[\protect\citeauthoryear{{Luan} \& {Goldreich}}{{Luan} \&
  {Goldreich}}{2014}]{Luan2014}
{Luan} J.,  {Goldreich} P.,  2014, \mn@doi [\apjl]
  {10.1088/2041-8205/785/2/L26}, \href
  {https://ui.adsabs.harvard.edu/abs/2014ApJ...785L..26L} {785, L26}

\bibitem[\protect\citeauthoryear{{Lyubarsky}}{{Lyubarsky}}{2014}]{Lyubarsky14}
{Lyubarsky} Y.,  2014, \mn@doi [\mnras] {10.1093/mnrasl/slu046}, \href
  {https://ui.adsabs.harvard.edu/abs/2014MNRAS.442L...9L} {442, L9}

\bibitem[\protect\citeauthoryear{{Main}, {Hilmarsson}, {Marthi}, {Spitler},
  {Wharton}, {Bethapudi}, {Li}  \& {Lin}}{{Main} et~al.}{2022}]{Main2022}
{Main} R.~A.,  {Hilmarsson} G.~H.,  {Marthi} V.~R.,  {Spitler} L.~G.,
  {Wharton} R.~S.,  {Bethapudi} S.,  {Li} D.~Z.,   {Lin} H.~H.,  2022, \mn@doi
  [\mnras] {10.1093/mnras/stab3218}, \href
  {https://ui.adsabs.harvard.edu/abs/2022MNRAS.509.3172M} {509, 3172}

\bibitem[\protect\citeauthoryear{{Marcote} et~al.,}{{Marcote}
  et~al.}{2017}]{Marcote2017}
{Marcote} B.,  et~al., 2017, \mn@doi [\apjl] {10.3847/2041-8213/834/2/L8},
  \href {https://ui.adsabs.harvard.edu/abs/2017ApJ...834L...8M} {834, L8}

\bibitem[\protect\citeauthoryear{{Margalit}, {Metzger}  \& {Sironi}}{{Margalit}
  et~al.}{2020}]{Margalit+20}
{Margalit} B.,  {Metzger} B.~D.,   {Sironi} L.,  2020, \mn@doi [\mnras]
  {10.1093/mnras/staa1036}, \href
  {https://ui.adsabs.harvard.edu/abs/2020MNRAS.494.4627M} {494, 4627}

\bibitem[\protect\citeauthoryear{{Masui} et~al.,}{{Masui}
  et~al.}{2015}]{Masui+15}
{Masui} K.,  et~al., 2015, \mn@doi [\nat] {10.1038/nature15769}, \href
  {https://ui.adsabs.harvard.edu/abs/2015Natur.528..523M} {528, 523}

\bibitem[\protect\citeauthoryear{{Max}, {Arons}  \& {Langdon}}{{Max}
  et~al.}{1974}]{Max1974}
{Max} C.~E.,  {Arons} J.,   {Langdon} A.~B.,  1974, \mn@doi [\prl]
  {10.1103/PhysRevLett.33.209}, \href
  {https://ui.adsabs.harvard.edu/abs/1974PhRvL..33..209M} {33, 209}

\bibitem[\protect\citeauthoryear{{Metzger}, {Berger}  \& {Margalit}}{{Metzger}
  et~al.}{2017}]{Metzger+17}
{Metzger} B.~D.,  {Berger} E.,   {Margalit} B.,  2017, \mn@doi [\apj]
  {10.3847/1538-4357/aa633d}, \href
  {https://ui.adsabs.harvard.edu/abs/2017ApJ...841...14M} {841, 14}

\bibitem[\protect\citeauthoryear{{Metzger}, {Margalit}  \& {Sironi}}{{Metzger}
  et~al.}{2019}]{Metzger+19}
{Metzger} B.~D.,  {Margalit} B.,   {Sironi} L.,  2019, \mn@doi [\mnras]
  {10.1093/mnras/stz700}, \href
  {https://ui.adsabs.harvard.edu/abs/2019MNRAS.485.4091M} {485, 4091}

\bibitem[\protect\citeauthoryear{{Michilli} et~al.,}{{Michilli}
  et~al.}{2018}]{Michilli+18}
{Michilli} D.,  et~al., 2018, \mn@doi [\nat] {10.1038/nature25149}, \href
  {https://ui.adsabs.harvard.edu/abs/2018Natur.553..182M} {553, 182}

\bibitem[\protect\citeauthoryear{{Narayan}}{{Narayan}}{1992}]{Narayan1992}
{Narayan} R.,  1992, \mn@doi [Philosophical Transactions of the Royal Society
  of London Series A] {10.1098/rsta.1992.0090}, \href
  {https://ui.adsabs.harvard.edu/abs/1992RSPTA.341..151N} {341, 151}

\bibitem[\protect\citeauthoryear{{Nimmo} et~al.,}{{Nimmo}
  et~al.}{2021}]{Nimmo2021}
{Nimmo} K.,  et~al., 2021, \mn@doi [Nature Astronomy]
  {10.1038/s41550-021-01321-3}, \href
  {https://ui.adsabs.harvard.edu/abs/2021NatAs...5..594N} {5, 594}

\bibitem[\protect\citeauthoryear{{Nimmo} et~al.,}{{Nimmo}
  et~al.}{2022}]{Nimmo2022}
{Nimmo} K.,  et~al., 2022, \mn@doi [Nature Astronomy]
  {10.1038/s41550-021-01569-9}, \href
  {https://ui.adsabs.harvard.edu/abs/2022NatAs...6..393N} {6, 393}

\bibitem[\protect\citeauthoryear{{Niu} et~al.,}{{Niu} et~al.}{2022}]{Niu2022}
{Niu} C.~H.,  et~al., 2022, \mn@doi [\nat] {10.1038/s41586-022-04755-5}, \href
  {https://ui.adsabs.harvard.edu/abs/2022Natur.606..873N} {606, 873}

\bibitem[\protect\citeauthoryear{{Ocker}, {Cordes}  \& {Chatterjee}}{{Ocker}
  et~al.}{2021}]{2021ApJ...911..102O}
{Ocker} S.~K.,  {Cordes} J.~M.,   {Chatterjee} S.,  2021, \mn@doi [\apj]
  {10.3847/1538-4357/abeb6e}, \href
  {https://ui.adsabs.harvard.edu/abs/2021ApJ...911..102O} {911, 102}

\bibitem[\protect\citeauthoryear{{Ocker} et~al.,}{{Ocker}
  et~al.}{2022}]{Ocker2022b}
{Ocker} S.~K.,  et~al., 2022, \mn@doi [\apj] {10.3847/1538-4357/ac6504}, \href
  {https://ui.adsabs.harvard.edu/abs/2022ApJ...931...87O} {931, 87}

\bibitem[\protect\citeauthoryear{{Pastor-Marazuela} et~al.,}{{Pastor-Marazuela}
  et~al.}{2021}]{2021Natur.596..505P}
{Pastor-Marazuela} I.,  et~al., 2021, \mn@doi [\nat]
  {10.1038/s41586-021-03724-8}, \href
  {https://ui.adsabs.harvard.edu/abs/2021Natur.596..505P} {596, 505}

\bibitem[\protect\citeauthoryear{{Pleunis} et~al.,}{{Pleunis}
  et~al.}{2021a}]{Pleunis2021a}
{Pleunis} Z.,  et~al., 2021a, \mn@doi [\apjl] {10.3847/2041-8213/abec72}, \href
  {https://ui.adsabs.harvard.edu/abs/2021ApJ...911L...3P} {911, L3}

\bibitem[\protect\citeauthoryear{{Pleunis} et~al.,}{{Pleunis}
  et~al.}{2021b}]{Pleunis2021b}
{Pleunis} Z.,  et~al., 2021b, \mn@doi [\apj] {10.3847/1538-4357/ac33ac}, \href
  {https://ui.adsabs.harvard.edu/abs/2021ApJ...923....1P} {923, 1}

\bibitem[\protect\citeauthoryear{{Puls}, {Vink}  \& {Najarro}}{{Puls}
  et~al.}{2008}]{Puls2008}
{Puls} J.,  {Vink} J.~S.,   {Najarro} F.,  2008, \mn@doi [\aapr]
  {10.1007/s00159-008-0015-8}, \href
  {https://ui.adsabs.harvard.edu/abs/2008A&ARv..16..209P} {16, 209}

\bibitem[\protect\citeauthoryear{{Qu}, {Kumar}  \& {Zhang}}{{Qu}
  et~al.}{2022}]{Qu2022}
{Qu} Y.,  {Kumar} P.,   {Zhang} B.,  2022, \mn@doi [\mnras]
  {10.1093/mnras/stac1910}, \href
  {https://ui.adsabs.harvard.edu/abs/2022MNRAS.515.2020Q} {515, 2020}

\bibitem[\protect\citeauthoryear{{Rybicki} \& {Lightman}}{{Rybicki} \&
  {Lightman}}{1979}]{Rybicki79}
{Rybicki} G.~B.,  {Lightman} A.~P.,  1979, {Radiative processes in
  astrophysics}

\bibitem[\protect\citeauthoryear{{Sammons} et~al.,}{{Sammons}
  et~al.}{2023}]{Sammons2023}
{Sammons} M.~W.,  et~al., 2023, \mn@doi [arXiv e-prints]
  {10.48550/arXiv.2305.11477}, \href
  {https://ui.adsabs.harvard.edu/abs/2023arXiv230511477S} {p. arXiv:2305.11477}

\bibitem[\protect\citeauthoryear{{Snelders} et~al.,}{{Snelders}
  et~al.}{2023}]{Snelders2023}
{Snelders} M.~P.,  et~al., 2023, \mn@doi [arXiv e-prints]
  {10.48550/arXiv.2307.02303}, \href
  {https://ui.adsabs.harvard.edu/abs/2023arXiv230702303S} {p. arXiv:2307.02303}

\bibitem[\protect\citeauthoryear{{Sobacchi}, {Lyubarsky}, {Beloborodov}  \&
  {Sironi}}{{Sobacchi} et~al.}{2021}]{sobacchi21_modulational_instability}
{Sobacchi} E.,  {Lyubarsky} Y.,  {Beloborodov} A.~M.,   {Sironi} L.,  2021,
  \mn@doi [\mnras] {10.1093/mnras/staa3248}, \href
  {https://ui.adsabs.harvard.edu/abs/2021MNRAS.500..272S} {500, 272}

\bibitem[\protect\citeauthoryear{{Sutton}}{{Sutton}}{1971}]{1971MNRAS.155...51S}
{Sutton} J.~M.,  1971, \mn@doi [\mnras] {10.1093/mnras/155.1.51}, \href
  {https://ui.adsabs.harvard.edu/abs/1971MNRAS.155...51S} {155, 51}

\bibitem[\protect\citeauthoryear{{The Chime/Frb Collaboration} Andersen
  et~al.,}{{The Chime/Frb Collaboration} et~al.}{2020}]{CHIME2020}
{The Chime/Frb Collaboration} Andersen B.~C.,  et~al., 2020, \mn@doi [\nat]
  {10.1038/s41586-020-2863-y}, \href
  {https://ui.adsabs.harvard.edu/abs/2020Natur.587...54T} {587, 54}

\bibitem[\protect\citeauthoryear{{Wang}, {Zhang}, {Chen}  \& {Xu}}{{Wang}
  et~al.}{2019}]{wang2019}
{Wang} W.,  {Zhang} B.,  {Chen} X.,   {Xu} R.,  2019, \mn@doi [\apjl]
  {10.3847/2041-8213/ab1aab}, \href
  {https://ui.adsabs.harvard.edu/abs/2019ApJ...876L..15W} {876, L15}

\bibitem[\protect\citeauthoryear{{Wang}, {Xu}  \& {Chen}}{{Wang}
  et~al.}{2020}]{wang+20}
{Wang} W.-Y.,  {Xu} R.,   {Chen} X.,  2020, \mn@doi [\apj]
  {10.3847/1538-4357/aba268}, \href
  {https://ui.adsabs.harvard.edu/abs/2020ApJ...899..109W} {899, 109}

\bibitem[\protect\citeauthoryear{{Zhang}}{{Zhang}}{2020}]{zhang-review2020}
{Zhang} B.,  2020, \mn@doi [\nat] {10.1038/s41586-020-2828-1}, \href
  {https://ui.adsabs.harvard.edu/abs/2020Natur.587...45Z} {587, 45}

\bibitem[\protect\citeauthoryear{{Zhang} et~al.,}{{Zhang}
  et~al.}{2023}]{Zhang2023}
{Zhang} Y.-K.,  et~al., 2023, \mn@doi [arXiv e-prints]
  {10.48550/arXiv.2304.14665}, \href
  {https://ui.adsabs.harvard.edu/abs/2023arXiv230414665Z} {p. arXiv:2304.14665}

\makeatother
\end{thebibliography}
\bibliographystyle{mnras}

\bigskip
\bigskip

\appendix
\section{Scintillation amplitudes for incoherent and coherent extended sources}
\label{sec:extendedsources}
Consider an extended source of transverse size $R_s$, and a turbulent plasma screen that lies in the host galaxy of the source. In this case $\ell_\pi^{\rm so} = \ell_\pi$. Moreover, let us consider that $R_s > \ell_\pi$. We break up the source into small segments of sizes slightly smaller than $\ell_\pi$ and add up their contributions to calculate the observed wave amplitude. The total number of segments is $N \sim (R_s/\ell_\pi)^2$. The wave amplitude
at the observer location from the i-th segment, after it has undergone scattering by the 
plasma screen, is given below by the Fresnel-Kirchoff integral over the screen surface,
\begin{equation}
	\label{eq:Ai}
  A_i(\nu) \propto \exp[ i\phi_s^\nu({\bf y_i})] \int dx^2 \exp\left[ { i({\bf x - y_i})^2 \over R_F^2} + i\phi_p^\nu({\bf x}) \right],
\end{equation}
where $\phi_s^\nu({\bf y_i})$ is the phase of the wave at the source location 
${\bf y_i}$, 
and $\phi_p^\nu({\bf x})$ is the phase shift suffered by the wave as it crosses the scattering screen at location ${\bf x}$.  The scintillation bandwidth
($\delta\nu_{\rm sc}$) is the frequency interval over which the phase term in equation (\ref{eq:Ai}) changes
by $\sim\pi$. Thus, $\delta\nu_{\rm sc}/\nu \sim (\ell_\pi/R_F)^2$. We note that the scintillation bandwidth is the same for all segments of the extended source.

The contribution to the total wave amplitude at the observer location from the
i-th segment of the source can be written as, 
$A_i(\nu) \sim {\sc Re}\left\{\delta A^\nu \exp[i\bar\phi_p^\nu({\bf y_i})]\right\}\exp[ 
i\phi_s^\nu({\bf y_i})]$; where $\delta A^\nu$ is the observed wave amplitude in the absence 
of the scattering screen, and $\bar\phi_p^\nu({\bf y_i})$ 
reflects how the amplitude is modified when the wave scattered by different parts 
of the screen interfere at the observer location. 
The phase $\bar\phi_p^\nu({\bf y_i})$ is a random variable with 
value between 0 and $\pi$, and thus the observed wave amplitude for the i-th patch 
of the source lies between $-\delta A^\nu$ and $\delta A^\nu$. The observed wave flux 
from the entire source in the observed frequency band $\delta\nu_{\rm ob}$ of the detector is:
\begin{equation}
   f(\nu) = \sum_{i,j} \int_\nu^{\nu+\delta\nu_{\rm ob}} d\nu\, A_i(\nu)A_j^*(\nu)
   \label{flux1}
\end{equation}

For a partially coherent source, wave phases at two different points in the source
change with frequency in a correlated way within a frequency band of width 
$\delta\nu_{\rm co}^s$ -- defined as the scintillation bandwidth of the 
source\footnote{In general, 
$\delta\nu_{\rm co}^s$ depends on the separation between the two points in the source. 
However, we are ignoring that here to keep the discussion simple.}. 
The sum of all off-diagonal terms in the above equation, for a partially coherent
source, are smaller than the sum of diagonal terms by a factor 
$(\delta\nu_{\rm ob}/\delta\nu_{\rm co}^s)^{1/2}$ when $\delta\nu_{\rm co}^s < \delta\nu_{\rm ob}$. The off
diagonal terms vanish in the limit $\delta\nu_{\rm co}^s\rightarrow 0$ for a
completely incoherent source for which $\bar\phi_p^\nu({\bf y_i})$ and 
$\bar\phi_p^\nu({\bf y_j})$ are uncorrelated. The observed flux
in this limit reduces to 
\begin{equation}
   f(\nu) = \sum_{i} \int_\nu^{\nu+\delta\nu_{\rm ob}} d\nu\, |A_i(\nu)|^2.
    \label{flux2}
\end{equation}
The frequency dependence of the observed flux for a small source (size less than 
$\ell_\pi$), with smoothly varying intrinsic spectrum, that is subject to strong 
scintillation is a stochastic function with scintillation bandwidth 
$\delta\nu_{\rm sc} \sim (\ell_\pi/R_F)^2\nu \sim 1/\delta t_s$. The scintillation bandwidth is the same for all different patches of the source, however, the location
of the peaks and troughs in the spectrum of an individual patch do depend on it's location (${\bf y_i}$).
Thus, each term of the series in equation (\ref{flux2}) fluctuates with frequency 
with mean separation between peaks being $\delta\nu_{\rm sc}$.
Therefore, an extended source has the same coherence bandwidth for scintillation as a point source, however,
the amplitude of flux variation is reduced by a factor $R_s/\ell_\pi$ in comparison to a
point source. This result applies only to an incoherent source where different segments of the source are completely uncorrelated.

For a source that is coherent across the entire region of size $R_s$, the sum of 
off-diagonal terms in equation \ref{flux1} is comparable to the sum of diagonal terms, 
and the RMS value of the flux is $N (\delta A^\nu)^2\delta\nu_{\rm ob}$. The amplitude of flux variation with frequency, in this case, 
is of the same order as the RMS value since the off-diagonal and diagonal terms
have similar magnitudes and the sign of the off-diagonal terms is random. Thus, 
$\delta f/f \sim 1$ for an extended coherent source of size much larger than $\ell_\pi$.
We note, however, that it is hard, if not impossible, to maintain coherence over
length scales larger than the wavelength of the wave. And thus sources of size
larger than $\ell_\pi$ are highly unlikely to be coherent in the sense we have discussed here.

\end{document}